\documentclass[useAMS,usenatbib]{mn2e}
\usepackage{graphicx}

\newcommand{\elem}[1]{{\scriptsize~{#1}}}

\newcommand{\phot}{~photon~cm$^{-2}$~s$^{-1}$~sr$^{-1}$}

\newcommand{\Xenia}{{\it Xenia} }
\usepackage{color}

\title[Expected properties of the Two-Point Autocorrelation Function of the IGM]{Expected properties of the Two-Point Autocorrelation Function of the IGM}
\author[E. Ursino et al.]{Ursino, E.$^{1}$\thanks{E-mail: ursino@fis.uniroma3.it}, Branchini, E.$^{1}$, Galeazzi, M.$^2$, Marulli, F.$^3$, Moscardini, L.$^{3,4}$, Piro, L.$^5$, 
	\newauthor Roncarelli, M.$^6$, and Takei, Y.$^7$\\
$^1$ Dipartimento di Fisica, Universit\'a di Roma TRE, Via della Vasca Navale 84, I-00146, Roma, Italy\\
$^2$ Physics Department, University of Miami, Coral Gables, FL 33155\\
$^3$ Dipartimento di Astronomia, Universit\'a degli Studi di Bologna, Via Ranzani 1, I-40127, Bologna, Italy\\
$^4$ INFN, Sezione di Bologna, viale Berti Pichat 6/2, I-40127 Bologna, Italy\\
$^5$ INAF-Istituto di Astrofisica Spaziale Fisica Cosmica, Via del Fosso del Cavaliere 100, I-00133 Roma, Italy\\
$^6$ Centre d'Etude Spatiale des Rayonnements (CESR), 9, av. du Colonel Roche - B.P. 44346, 31028, Toulouse Cedex 4, France\\
$^7$ Institute of Space and Astronautical Science (ISAS), JAXA, 3-1-1 Yoshinodai, Chuo-ku, Sagamihara, Kanagawa 252-5210 Japan\\}

\begin{document}

\date{Accepted . Received ; }

\pagerange{\pageref{firstpage}--\pageref{lastpage}} \pubyear{2010}

\maketitle

\label{firstpage}

\begin{abstract}
Recent analyses of the fluctuations of the soft Diffuse X-ray 
Background (DXB) have provided indirect detection of a component 
consistent with the elusive Warm Hot Intergalactic Medium (WHIM). In 
this work we use theoretical predictions obtained from hydrodynamical 
simulations to investigate the angular correlation properties of the 
WHIM in emission and assess the possibility of indirect detection with 
next-generation X-ray missions. Our results indicate that 
the angular correlation signal of the WHIM is generally weak but 
dominates the angular correlation function of the DXB outside 
virialized regions. Its indirect detection is possible but requires 
rather long exposure times
[0.1-1]~Ms, large ($\sim 1^{\circ} \times 1^{\circ}$) 
fields of view and accurate subtraction of isotropic fore/background 
contributions, mostly contributed by Galactic emission.
The angular correlation function of the WHIM, which turns out to be 
positive for $\theta < 5'$ provides limited information on its 
spatial distribution. A satisfactory characterization of the WHIM in 
3D can be obtained through spatially resolved spectroscopy. 1~Ms long 
exposures with next generation detectors will allow to detect $\sim 
400$ O\elem{VII}+O\elem{VIII} X-ray emission systems that could be 
used to trace the spatial distribution of the WHIM. We predict that 
these observations will allow to estimate the WHIM correlation function with
high statistical significance out to $\sim 10$~Mpc~h$^{-1}$ and characterize 
its dynamical state through the analysis of redshift-space distortions. 
The detectable WHIM, which is typically associated with the outskirts of virialized regions
rather than the filaments has a non-zero correlation function with slope 
$\gamma=-1.7\pm0.1$ and correlation length $r_0=4.0\pm0.1$~Mpc~h$^{-1}$ 
in the range $r=[4.5,12]$~Mpc~h$^{-1}$. Redshift space distances can be 
measured to assess the dynamical properties of the gas, that we predict 
to be typically infalling onto large virialized structures.
\end{abstract}


\section{Introduction}

The observed abundance of baryons in the local Universe does not agree 
with the high redshift observations \citep*{Fukugita98}. The Lyman alpha 
forest at $z=2$, the results from the WMAP experiment, and the 
predictions of the standard nucleosynthesis model are all consistent 
with a baryonic density $\Omega_B\simeq0.04$ \citep{Rauch98, 
Weinberg97, BurTyt98, Kirkman03, Bennett03, Spergel07, Komatsu09}. 
At variance with these results, the baryon mass in stars, galaxies, 
and clusters in the local Universe seems to be $2-4$ times lower.

Large-scale cosmological hydrodynamic simulations 
\citep*{CenOst99, Croft01, Yoshikawa03, Chen03, Borgani04, CenOst06, 
Shen010, Smith10, Tornatore10} 
allow to follow the evolution of baryons from 
high redshift to present days. A common feature of these simulations 
is that $40-50~\%$ of the baryons in the Universe at $z\sim 0$ are in the 
form of highly ionized gas with temperature in the range of 
$10^5-10^7$~K. This gas is characterized by a spatial distribution that 
closely resembles that of the underlying cosmic web of the Dark Matter, 
and is commonly indicated as the Warm-Hot Intergalactic Medium (WHIM). 
According to thermal emission models for gas in collisional 
equilibrium \citep{Raymond77} and assuming an average metallicity of 
$0.1$~Z$_\odot$, this highly ionized gas emits mainly in the soft 
X-ray ([0.1-1]~keV) and in the Far and Extreme UltraViolet 
(FUV, [0.01-0.1]~keV) energy bands. However, the expected WHIM 
contribution to the total surface brightness (SB) in these bands is 
expected to be weak. The comparison between the predicted emission of the WHIM 
and the measured Diffuse X-Ray Background (DXB), for example, shows 
that the WHIM contribution to the total SB is in the order of $10\%$ 
\citep*{Phillips01, Ursino06, Ursino10}. Highly ionized metals in the 
warm-hot gas are responsible for line emission or absorption features 
whose strength is also expected to be rather weak \citep*{Kravtsov02, 
Chen03, Klypin03, Viel03, Yoshikawa03, Yoshikawa04, CenFang06, 
Branchini09}. 

Observational data confirm the difficulty to detect 
WHIM lines, both in absorption and emission. A clear detection of the 
WHIM comes from the absorption lines of ions like O\elem{VI}, 
N\elem{V}, C\elem{III}, C\elem{IV}, Si\elem{III}, Si\elem{IV}, and 
Fe\elem{III} in the FUV spectra of bright, distant sources. Roughly a 
hundred FUV absorbers have been detected in the spectra of
$\sim30$ AGN \citep*{Danforth05, Danforth08, Thom08, Tripp08}. 
Their abundance turned out to be in agreement with theoretical 
predictions  \citep{CenFang06}. However, these 
absorbers trace only the warm gas (T$<10^6$~K) whereas the bulk of the
WHIM is expected to be at higher temperatures.
In the soft X-ray band the search for the WHIM has led to a few detections 
\citep*{Fang02,Nicastro05, Fang07, Buote09, Fang10,Williams010} whose 
statistical significance is still under debate \citep*{Kaastra06, Rasmussen07}. In 
emission, the only  claim for a line detection so far 
is likely to be associated with a
gas filament connecting the two galaxy clusters 
A222 and A223 \citep{Werner08}. However, the overdensity ($\delta \simeq150$)
and the temperature  (T$\sim1\times10^7$~K) of this gas
are only marginally consistent with the WHIM definition.
In addition, \citet*{Zappacosta02, Mannucci07} observed a soft X-ray excess associated to galaxy 
concentration and possibly consistent with WHIM emission. 

Finally, an indirect evidence for the WHIM  comes from the 
analysis of the Diffuse DXB in the soft band [0.4-1.0]~keV,
which is known to be mainly  contributed by unresolved AGN, starburst galaxies 
and local diffuse 
emission \citep*{Snowden97, Mush00, McCammon02, Galeazzi07, Henley08, 
Gupta09a, Henley10}. 
These sources account for all but $10\%$ to $20\%$ of the DXB 
\citep{Hickox07} matching the theoretical expectations 
for the WHIM emission in this band \citep{Roncarelli06}. 

The WHIM is also expected to provide some contribution, perhaps with 
characteristic features, to the angular correlation properties of the 
DXB fluctuations. Such contribution, if detected, would provide a 
further, indirect evidence for the WHIM. Indeed, a recent angular 
correlation analysis of the unresolved DXB measured by {\it XMM} 
\citep*{Galeazzi09} has revealed the presence of a component that 
has been identified with the WHIM.
Future X-ray missions are expected to detect a large number of 
characteristic emission lines in the X-ray band through which one will be able to 
unambiguously detect the WHIM \citep{CenFang06} and probe its spatial distribution 
much more  densely than it could be 
possibly done in absorption, due to the paucity of the background 
sources \citep*{Viel03, Viel05}.
Therefore one expects that only 
emission studies in the X-ray band will be able not only to detect the 
WHIM and characterize its thermal state, but also to provide 
a tomography of a large fraction of the missing baryons in the local universe.
In particular, as we will show in this work, given the characteristic of next-generation
X-ray detectors, the estimate of the spatial two-point correlation function of the 
WHIM line-emitting regions will allow to characterize the spatial 
correlation properties of the WHIM and its dynamical state. 

In this work we investigate the possibility of characterizing the 
angular and spatial distribution of the WHIM using next-generation  X-ray satellites.
More specifically, we aim at two different but related goals. First we 
investigate the possibility of finding the signature of the WHIM 
in the angular correlation function of the unresolved DXB.
Second, we assess the possibility of tracing the 
spatial distribution and the dynamical state of 
the WHIM by detecting its characteristic emission lines and 
measuring the spatial two-point correlation function of the line-emitting regions. 
The two ingredients required for this goal are:
a theoretical WHIM model to rely upon and a reference experimental 
setup. 
For the WHIM model we assume the same one as of \citet{Ursino10} 
and \citet{Branchini09}, based on the \citet{Borgani04} 
hydrodynamical simulation. As a reference 
experimental setup we consider the recently proposed \Xenia
\footnote{http://sms.msfc.nasa.gov/xenia/} 
X-ray satellite \citep{Hartmann09}. The reason for considering 
this particular mission is that it is designed to 
carry two instruments well suited for our purposes: the High Angular Resolution Imager (HARI) 
and the Cryogenic Imaging Spectrometer (CRIS). The HARI is a CCD or CMOS 
based detector with a field of view (f.o.v) shaped as a circle of diameter $1.4^\circ$,
angular resolution of $15''$ for on-axis objects, and energy 
resolution of $50-150$~eV FWHM (in the range $0.3-5.0$~keV). 
The CRIS is a spatially 
resolved spectrometer based on Transition Edge Sensor technology.
The imaging area is $0.9^\circ\times0.9^\circ$,  angular 
resolution is $\sim 2.5'$, and the energy resolution is in the 
range [1-3] eV. The X-ray 
telescope has an effective area of 1000~cm$^2$ at 1~keV. 
The HARI instrument is expected to provide the possibility to
study the angular correlation 
properties of the DXB and isolate the WHIM contribution thank to its
large f.o.v. and angular resolution, smaller than the typical angular size
of the WHIM emitting element \citep{Ursino06}.
The CRIS instrument will allow to detect the 
WHIM emission lines and identify the spatial position of the emitting gas
thanks to the good angular and energy resolution.

The plan of the paper is as follows. In Section \ref{Generation} we 
introduce the WHIM model  and the procedure we adopted to generate 
the simulated X-ray maps and spectra that will be considered for the 
correlation analysis. In Section \ref{2DACF} we describe the estimator 
used to compute the angular two-point correlation function and estimate 
the contribution of the WHIM to the DXB angular correlation signal in 
the case of a simulated observation with \Xenia. In Section \ref{3DACF} 
we measure the spatial two-point correlation function in the simulated 
maps and assess the possibility of characterizing the spatial 
distribution and dynamical state of the WHIM in the case of a \Xenia 
observation.In Section \ref{conclusions} we summarize and discuss the 
results of our work.

\section{Simulated datasets}
\label{Generation}
In this section we briefly describe the hydrodynamical simulation of 
\citet{Borgani04}, the WHIM model built upon these 
 \citep{Ursino10, Branchini09} and the mock X-ray SB maps and spectra obtained
from this model taking into account the response functions of the 
HARI and CRIS detectors. 

\subsection{The Hydrodynamical Model of the WHIM}
\label{Model}

The WHIM model considered in this work has been obtained from the 
hydrodynamical simulation of \citet{Borgani04}. This simulation 
assumes a flat ($\Omega_m=0.3$, $\Omega_\Lambda=0.7$) $\Lambda$CDM 
model with a nonvanishing cosmological constant, with $h=0.7$, 
$\Omega_b=0.04$, and $\sigma_8=0.8$. The simulation was performed 
using the TREESPH code GADGET-2 \citep*{Springel01, Springel05} with a 
Plummer equivalent softening, fixed in comoving units, 
$\epsilon_{Pl}=7.5$~$h^{-1}~kpc$ at $z<2$, and fixed in physical units 
at $z>2$. The simulation follows the evolution of $480^3$ dark matter 
(DM) particles and $480^3$ baryonic gas particles from redshift 
$z=49$ to $z=0$. The simulation box has a side of 192~$h^{-1}$~Mpc. DM 
and gas particles have masses 
$m_{DM}=4.62\times10^9$~$h^{-1}$M$_\odot$ and 
$m_{gas}=6.93\times10^8$~$h^{-1}$M$_\odot$, respectively. Radiative 
cooling was performed under the assumption of  optically thin gas 
with pristine composition ($76\%$ hydrogen and $24\%$ helium) and in 
collisional equilibrium. The simulation includes a time-dependent 
photoionizing UV background that 
reionizes the Universe at $z\simeq6$ \citep{Haardt96}. A two-phase 
model \citep{Springel03} and a Salpeter initial mass function 
\citep{Salpeter55} were assumed in the simplified star formation model. 
Galactic winds with typical velocity of about 350~km~s$^{-1}$ were assumed to carry energy, 
mass, and metals produced by stars in the IGM \citep{Springel03}. The 
simulation output consists of 102 boxes, equally spaced in logarithmic redshift
scale between $z=9$ and $z=0$. This simulation correctly reproduces 
the observational properties of rich X-ray emitting clusters 
\citep{Borgani04} but overestimates the emission from smaller 
structures that can be identified with galaxy groups. 
This overestimate probably originates from 
having assumed primordial composition in the cooling 
function, instead of updating the cooling rate according to the 
evolving metallicity of the gas \citep{Bertone10}. 
However, this effect is not expected to 
affect significantly the emission properties of the WHIM gas 
in regions of lower density, outside virialized structures, 
 characterized by an average metallicity well below solar
 \citep{Shen010}
The fact that our WHIM model does not exceed the DXB 
constraint \citep{Roncarelli06} corroborates this assumption. 
In fact, the overestimate of the X-ray emission from galaxy groups 
artificially reduces the contribution of the WHIM to the DXB.
In this respect, all theoretical prediction on the 
WHIM detectability presented in this work are to be 
regarded as conservative.

Metal diffusion is described by particles that actually undergo metal 
enrichment by star formation, and not by transfer of metals between 
neighboring particles \citep*{Wiersma09a}. As a result, the spatial 
distribution of metals may result unrealistically inhomogeneous, in 
particular for the gas in lower density regions (like the WHIM), far 
away from star forming regions. To circumvent  this potential problem, we 
did not use the metallicity calculated self-consistently within the 
simulation. Instead, we assigned a metallicity to all gas particle in each simulation output 
by matching the metallicity-density relation
in the simulation of \citet{CenOst99b}. In practice, at a given output, 
we consider each gas particle with density $\rho$,
and assign a gas metallicity according to the 
probability function $P(Z(\rho,z))$ built upon the simulation outputs of 
\citet{CenOst99}, as fully described in \citet{Ursino10}. It is reassuring 
that, with this metallicity prescription, the WHIM model correctly 
reproduces the observed abundances of O\elem{VI} absorbers
and satisfies the current 
constraints for the O\elem{VII} lines  \citep{Branchini09}.

Since we are interested in modeling X-ray observations of the WHIM, 
we focused our analysis on gas at temperatures higher than $10^5$~K. 
We adopted the definition of the WHIM as the gas with 
$10^5 < \textrm{T} < 10^7$~K and 
$\rho  \le1000 \langle\rho\rangle$, distributed along a network of 
filaments connecting virialized structures and in their outskirts.
For the purpose of characterizing the angular and spatial correlation 
properties of the WHIM, however, we find it more convenient to 
consider different classifications for the gas phases that we describe 
here for reference.

In \S~\ref{2DACF}, to study the angular correlation properties of the 
DXB, we find it more convenient to divide the gas into three phases:

\begin{itemize}
\item The WHIM, which coincides with the previous definition. 
\item The {\it non-}WHIM.
\item The {\it Total Gas}, which represents the sum of the two 
contributions, i.e. all gas with $\textrm{T}>10^5$~K. 
\end{itemize}

In \S~\ref{3DACF}, to study the spatial correlation properties of the 
line-emitting gas, we consider the following phases:
\begin{itemize} 
\item The {\it Total Gas}, the gas responsible for detectable line 
emission, irrespective of its thermal state or number density. 
\item The {\it Bright}-WHIM represents gas regions in which the WHIM 
contributes to the emission of detectable X-ray lines (i.e. regions 
where the WHIM contribution is above the detectability threshold, 
described in \S~\ref{generate_spectra}), regardless whether the non-WHIM 
contribution is stronger of the WHIM contribution or not, or whether the 
non-WHIM contribution is above the detectability threshold. 
\item The WHIM-{\it dominated} gas, represents gas regions in which 
the WHIM contribution to the line emission is larger than that of 
all other phases and the sum of the WHIM and non-WHIM contributions 
is above the detection threshold, even though the WHIM contribution 
alone is not above the detection threshold.
\end{itemize}
The {\it Total Gas} traces dense and bright regions (usually groups 
and clusters), as well as the less dense regions where the WHIM 
resides (outskirts and bright filaments). The {\it Bright}-WHIM gas 
could trace also the brightest regions if, model-wise, they are 
multiphase and have a WHIM contribution above detectability. The 
WHIM-{\it dominated} gas traces only the bright filaments and the 
outskirts of virialized structures, while in the cores of virialized 
structures the WHIM contributes to emission less than the hot and 
dense gas.

\subsection{Mock X-Ray emission maps}
\label{generate_maps}

To study the angular correlation properties of the WHIM and assess the possibility of 
its indirect detection we have generated a suite of mock SB maps in the soft 
X-ray band. Here we summarize the procedure to generate the maps and 
refer  the interested reader to
\citet{Ursino10} for a more detailed description.
As a first step we consider all gas particles within a light-cone, 
extracted from the hydrodynamical simulation. The light-cone is
obtained by stacking 
the simulation output at different epochs out to $z=2$
\citep{Roncarelli06}. X-ray maps are obtained by considering 
 the individual contribution of all the particles within the light-cone
to the SB. In particular, for every particle we 
obtain the X-ray spectrum from the APEC model of the
XSPEC~\footnote{http://heasarc.gsfc.nasa.gov/docs/xanadu/xspec/} 
package to simulate the emission from the plasma. 
Spectra are fully specified by the temperature, 
density, and metallicity of the particle. 
Here we focus on the energy range $0.380-0.650$~keV, where the WHIM contribution to the 
DXB is thought to be maximum. The lower  limit of 0.38~keV 
allows us to exclude the strong local C\elem{VI} line (0.367~keV) that 
would superimpose and dominate the weaker redshifted O\elem{VII} and 
O\elem{VIII} lines.  The upper limit allows us to include the O\elem{VII} 
triplet and the O\elem{VIII} line at redshift zero, as well as redshifted lines of 
heavier metals like Fe\elem{XVII} (0.826~keV), Ne\elem{IX} (0.922~keV), 
and Mg\elem{XI} (1.352~keV).
The use of the APEC model is justified under the assumption that
the gas is in collisional equilibrium. The goodness of this hypothesis is 
guaranteed by the fact that emission is proportional to the square of 
density. This is also valid for the WHIM contribution which, therefore, 
mostly comes from high 
density regions in which the gas is usually hotter and the ionization 
balance is set by collisions only. 

To account for the angular resolution of the detector we have 
binned the particles' contributions to the flux into square 
pixels. 
The flux in each pixel account for the contribution from all particles within the 
pixel itself plus that of particles in neighboring pixels 
weighted according to their smoothing kernel. All maps 
account for the effect of absorption due to the neutral hydrogen in 
our Galaxy. We modeled absorption according to \citet{Morrison83}, 
applying a typical value at high latitude of 
$\textrm{N}_\textrm{\scriptsize{H}}=1.8\times10^{20}$
~cm$^{-2}$ \citep{McCammon02}. 
Overall we have generated 10 simulated maps of the DXB in the energy 
range $380-650~$eV. We have binned the $1^\circ\times1^\circ$ f.o.v 
with $256\times256$ pixels in order to obtain an angular resolution of 
 $14^{"}\times14^{"}$ comparable to that expected for  the HARI
instrument. 
These maps constitute the database that we will analyze  
in  \S~\ref{2DACF}. 

\begin{figure}
\includegraphics[width=0.5\textwidth]{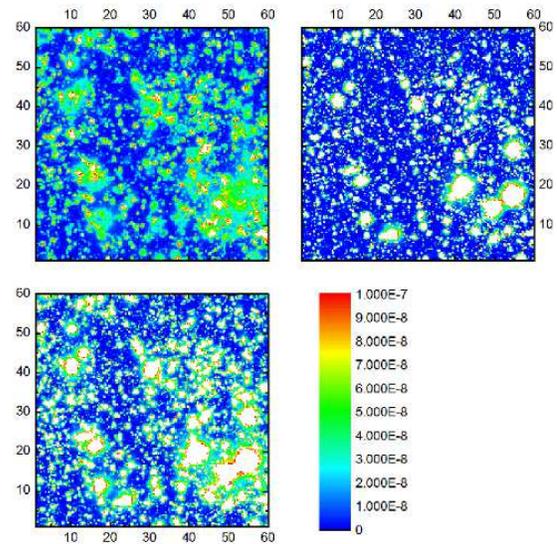}
\caption{Mock Surface brightness in the 
$380-650$~eV energy band, f.o.v. of
$1^\circ\times1^\circ$ and angular resolution 
$14^{"}\times14^{"}$. Contributions 
from the WHIM (\emph{top-left}) and {\it non}-WHIM 
(\emph{top-right}) phases are shown together with the 
 {\it Total Gas} signal (\emph{bottom-left}).
Color scale indicates different SB levels in  
units of photon~cm$^{-2}$~s$^{-1}$. 
\label{maps}}
\end{figure}

Fig.~\ref{maps} shows a  typical mock SB map
with a  f.o.v. of $1^\circ\times1^\circ$ and 
angular resolution of $14^{"}\times14^{"}$. 
The bottom panel shows the flux in the $0.380-0.650$~keV band 
contributed by all gas particles with $\textrm{T}>10^5$~K (i.e. the {\it Total gas}).
The top-left and top-right panels show the contribution of the WHIM and {\it non}-WHIM
phases, respectively.
These mock maps also include an isotropic
signal contributed by a
a Galactic foreground \citep{McCammon02} 
and a diffuse extragalactic background of unresolved AGN, that
we describe in details in \S~\ref{acf-xenia}.

In Tab.~\ref{table_components}  we have listed the mean SB 
contributed by the various components: WHIM, {\it non}-WHIM, 
Galactic foreground  and  unresolved AGN. 
While, as already pointed out, the WHIM contribution is consistent with observational 
constraints, the {\it non}-WHIM component exceeds these limits  \citep{Borgani04}. 
To quantify the mismatch is convenient to refer to the $0.65-1$~keV band 
in which observational  constraints have been derived by \citet{Hickox07}.
The observational value of $(1.0\pm0.2)\times10^{-12}$~ergs~cm$^{-2}$~s$^{-1}$~deg$^{-2}$
is reassuringly larger than the expected WHIM contribution
of 
$(3.6\pm0.3)\times10^{-13}$~ergs~cm$^{-2}$~s$^{-1}$~deg$^{-2}$. 
On the contrary the {\it non}-WHIM contribution of
$(8.7\pm0.5)\times10^{-12}$~ergs~cm$^{-2}$~s$^{-1}$~deg$^{-2}$ 
largely exceeds the observational constraints as a result of the 
fact that the simulation overestimates the
X-ray luminosity associated to galaxy groups, as anticipated in the 
previous section.

\begin{table}
\begin{center}
\caption{Contributions to the DXB from the different gas phases 
in the $380-650$~eV energy band. 
\label{table_components}}
\begin{tabular}{|c|c|}
\hline\hline
 & photon~cm$^{-2}$~s$^{-1}$~deg$^{-2}$ \\
\hline
WHIM$^a$      									&$(1.6\pm0.2)\times10^{-3}$  \\
{\it non}-WHIM$^{a}$		            &$(8.9\pm1.1)\times10^{-3}$  \\
Galactic Foreground$^b$    	&$(3.39\pm2.26)\times10^{-3}$  \\
AGN$^c$                   	&$(0.53\pm0.03)\times10^{-3}$  \\
\hline
\end{tabular}
\end{center}
\begin{flushleft}
$^a$ Mean values and standard deviations calculated from 10 maps.\\
$^b$ Model by \citet{McCammon02}, cosmic variance estimated from  
\citet{Gupta09b, Henley10}.\\
$^c$ Double power-law model (see section \ref{agn}), 
errors extrapolated from \cite{Moretti03}.
\end{flushleft}
\end{table}

\subsection{Mock 2D spectra}
\label{generate_spectra}

To  study the spatial correlation properties of the WHIM we have 
simulated data cubes representing a 2D-spectrum similar to that 
that will be taken by a spectrograph like CRIS.
For this purpose we have used our WHIM model to obtain 
a ``map'' consisting of a collection of 1D-spectra from contiguous 
pixels across a $5.5^\circ\times5.5^\circ$ area. 
Here we just summarize the main step of the 
procedure used to generate these  mock spectra and refer the interested 
reader to \citet{Takei10} for a more detailed description.
The possibility of performing a 3D correlation analysis of the 
WHIM relies on the possibility of detecting its characteristic 
X-ray emission lines. Detection estimates using our model have been 
obtained by \citet{Takei10} and will be adopted here.

To produce the spectra we adopt a procedure similar to that of the mock
SB maps except that here we consider larger pixels
($\sim1.3'\times1.3'$ corresponding to $\sim$ 50\% of the CRIS angular resolution)
and wider f.o.v.
($5.5^\circ\times5.5^\circ$,  roughly corresponding to a 
mosaic of 25 CRIS pointings).
The result is a collection of  $256\times256$ mock 1D-spectra, 
one for each pixel.
Furthermore, to account for the energy resolution of the instrument, 
we divide the particle distribution along each angular resolution 
element into radial bins of comoving length 3~Mpc~h$^{-1}$ (approximately
corresponding to $\Delta E\sim1$~eV).
We end up with $256\times256\times448\simeq3\times10^7$
mock resolution elements containing spectra convolved with the response 
matrix of the instrument that include the  effect of Galactic 
absorption, modeled as in \citet{Morrison83}, with a column density 
of $1.8\times 10^{20}$ ~cm$^{-2}$ \citep{McCammon02}.

To reduce the CPU time and memory requirement
we have  considered only gas particles out to 
$z=0.5$, corresponding to 
a maximum comoving depth of our correlation analysis of 
$\sim$ 1344~Mpc~h$^{-1}$.
In addition, since we are interested in the O\elem{VII} and O\elem{VIII} lines only, 
spectra in each bin have been generated using only particles that 
provide contribution in  the narrow energy range corresponding 
to that of the two redshifted lines.

To perform the 3D correlation analysis
we consider only those resolution elements in which the surface brightness of 
the O\elem{VIII} line is above $7\times10^{-2}$~\phot, a threshold that 
guarantees a 5-$\sigma$ detection in a 1~Ms observation with the CRIS
instrument \citep{Takei10}. 
We consider the O\elem{VIII} line only, rather than 
the simultaneous detection of O\elem{VII} and O\elem{VIII}. This 
choice is justified by the fact that  $\sim90\%$ of O\elem{VIII} 
brighter than our selection threshold is associated to a detectable 
O\elem{VII} line, as shown by \citet{Takei10} (Table~1 Model B2, 
$t_{exp}=1$~Ms). The  visual inspection of the 3D maps corroborates 
the remarkable spatial coincidence between the O\elem{VII}+O\elem{VIII} 
and the O\elem{VIII}-only line-emitting regions \citep{Takei10}.

In the geometry of our problem, the comoving volume of the resolution 
element increases with redshift. Since the typical size of line-emitting regions
is not expected to evolve significantly out to  $z=0.5$, the risk is of 
over- or under-sampling a line-emitting region, depending on its redshift.
To circumvent the problem we have grouped 
the spatially-contiguous resolution elements with detectable O\elem{VIII} 
lines into a single object, which we call {\it emitter},
and study the correlation of such emitters.
To assess the robustness of our results with respect to the grouping scheme,
we have implemented two different grouping procedures. 
In the first, we link all resolution elements having one 
side in common. In the second, we  group the 
elements that have one edge or one vertex in common.
We have verified that using either schemes does not significantly 
modify the results and therefore
we will only discuss the results obtained with the first, more conservative, grouping 
scheme. 
The outcome of the grouping procedure is a list of 
O\elem{VIII} emitters characterized by their total O\elem{VIII} line surface 
(obtained by summing over the grouped elements)  and spatial position
(defined as the SB-weighted centre of mass of the emitter).
As a final remark, we note that the 3~Mpc~h$^{-1}$ binning introduces a spurious 
periodicity along the line of sight which we eliminate by
randomizing the position of the emitter along  the line of sight, within 
the resolution element.

\section{ Angular correlation analysis of the  WHIM }
\label{2DACF}

In this section we address the problem of detecting the WHIM signature 
in the angular correlation signal of the DXB. We first describe the 
statistical tool used for the analysis, i.e. the two-point correlation 
function, then we apply it to the ideal case of an arbitrarily long 
exposure time and no contribution from AGN and Galactic foreground.
Then we include all contaminations to simulate a 
realistic observational setup.

\subsection{The angular correlation function}
\label{AcF-test}

As shown in Tab.~\ref{table_components}, the Galactic 
foreground and the {\it non}-WHIM gas dominate the
mean DXB signal in the $380-650$~eV energy range. The component
we are interested in, the WHIM, is largely subdominant (less than 10\%)
but its spatial distribution, from hydro simulations, is
markedly different from that
of the other components. Under such circumstances, which are reminiscent
to that of the CMB signal, the best option to infer the presence of the WHIM is 
by analyzing the two-point angular correlation function (ACF) 
of the DXB.
 
We characterize the angular correlation properties of the DXB 
by measuring the angular two-point correlation function $w(\theta)$
 of the signal in the mock maps. To measure $w(\theta)$ we use
 the  Landy-Szalay estimator \citep{Landy93} [LS from now on]:
\begin{equation}\label{acf_LS}
1+\hat{w}(\theta)=\frac{DD-2DR+RR}{RR} ,
\end{equation}
where $\theta$ is the angle between two pixels in the map and
\begin{equation}
DD(\theta)=\frac{\sum g_{ij} g_{kl}}{(\sum g_{ij})^2},
\end{equation}
\begin{equation}
DR(\theta)=\frac{\sum g_{ij} r_{kl}}{\sum g_{ij}\sum r_{kl}},
\end{equation}
\begin{equation}
RR(\theta)=\frac{\sum r_{ij} r_{kl}}{(\sum r_{ij})^2}.
\end{equation}
$DD$ represents the number of photon pairs, normalized to the total 
counts, obtained by summing over all photon counts $g_{ij}$ in two 
pixels $i$ and $j$ separated by $\theta$. $RR$ is the analogous 
quantity for a map in which the photons are uniformly distributed 
across the f.o.v (the Random sample). $DR$ represents the mixed counts 
contributed by pixels in the real and random dataset. We have verified 
that the LS estimator is equivalent to that used by other authors 
\citep{Ursino06, Galeazzi09}.

In this work we estimate the ACF of the DXB, the ACFs of its 
components and their cross-terms. 
For the simple case in which the DXB is the sum of two components, $A$ and $B$, 
with ACF $\omega_A$ and $\omega_B$ respectively, the following relation
holds for the  ACF of the DXB;

\begin{equation}\label{add_acf}
n_T^2w_T(\theta)=n_A^2w_A(\theta)+n_B^2w_B(\theta)+2\,n_An_Bw_{AB}(\theta),
\end{equation}
where $n_A$ and $n_B$ are the photon counts contributed by the two 
components, $n_T=n_A+n_B$, and $\omega_{AB}$ is the cross-correlation term.

Uncertainties in the ACF are estimated as the RMS scatter 
among the correlation functions measured among the ten, statistically independent, 
mock maps . 

\subsection{Ideal Case}
\label{acf-ideal}

Are the angular correlation properties of the WHIM significantly different from 
those of the other components? We address this question by computing the
ACF in the ideal case of an arbitrarily long observation with the HARI instrument, 
ignoring all additional components like the Galactic foreground (which 
is assumed to be perfectly subtracted) and unresolved sources like AGN 
(that are assumed to provide a negligible contribution, as we will 
show in Section \ref{agn}).

\begin{figure}
\includegraphics[width=0.5\textwidth]{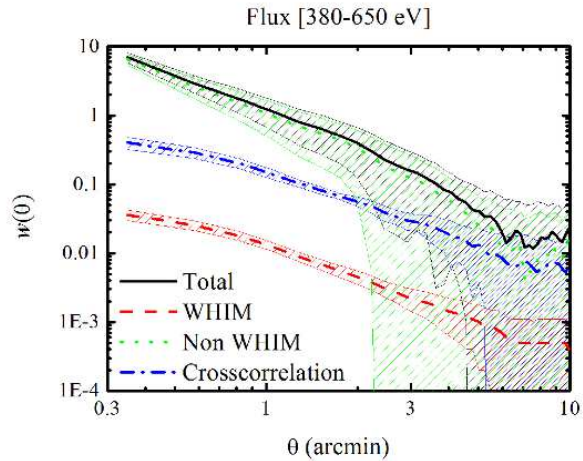}
\caption{ACFs of the WHIM  (\emph{dashed red}),   {\it non}-WHIM  (\emph{dotted 
green}), cross correlation term (\emph{dot-dashed blue}) and their sum
(\emph{solid black}). Amplitudes are normalized to the total flux. 
\label{xenia_flux_cross}}
\end{figure}

For the purpose of detecting  the WHIM signature, it is useful to 
compare the ACF of the flux map of the  WHIM and {\it non}-WHIM 
components. Fig.~\ref{xenia_flux_cross} compares the ACF of the WHIM 
(red, dashed line) with that of the {\it non}-WHIM (dotted, green). 
Lines represent the mean ACFs of the ten mock mapsvvand the band their 
rms scatter. As expected, the {\it non}-WHIM provides the dominant 
contribution to the total signal (solid, black line). Even the 
cross-correlation  signal (dot-dashed, blue line), estimated by 
subtracting the WHIM and {\it non}-WHIM contributions from the total 
ACF signal, is higher than that of the WHIM. However the WHIM 
autocorrelation signal is significantly above zero out to 
$\theta \simeq 6'$. The fact that the WHIM and {\it non}-WHIM ACFs 
have similar shapes is rather disappointing since it means that no 
characteristic feature in the ACF of the WHIM can be used to 
unambiguously infer its presence. On the other hand, the small 
amplitude of the WHIM correlation signal is less of a problem, 
considering that all ACFs in Fig.~\ref{xenia_flux_cross} are 
normalized to the total flux. Indeed, the amplitude of the 
{\it intrinsic} ACF of the WHIM, i.e. any term $w(\theta)$ in 
Eq.~\ref{add_acf}, is similar to that of the {\it non}-WHIM. The 
difference in amplitude reflects the different mean flux of the two 
phases: a suppression factor of $\left(\frac{n_\textrm{\tiny WHIM}}
{n_{non-\textrm{\tiny WHIM}}}\right)^2\sim 30$, with 
n$_\textrm{\tiny WHIM}$ and n$_{non-\textrm{\tiny WHIM}}$ the WHIM 
and the {\it non}-WHIM SB respectively, is expected between the ACF 
of the WHIM and that of the {\it non}-WHIM. This is encouraging and 
suggests that the WHIM correlation signal could be extracted from the 
total one, provided that the ACF is estimated in regions where the 
WHIM emission dominates over the {\it non}-WHIM one. Such regions 
are clearly seen in Fig.~\ref{maps} and show that WHIM emission is 
preferentially located at the outskirts of virialized structures 
whose central regions are, on the contrary, dominated by the 
{\it non}-WHIM component. 

These mock maps suggest that a possible strategy to extract the WHIM 
autocorrelation signal is to exclude from the correlation analysis the 
pixels associated to the large virialized structures in which the SB is usually larger.
To optimize the cleaning procedure 
one has to compromise between discarding pixels in which the signal 
is dominated by the {\it non}-WHIM component, and the need of using as 
many pixels as possible to increase the statistical significance of 
the signal. 
Our mock maps, in which we can pinpoint the 
contribution of the different components, allow us to calibrate this 
procedure. We have found that 
a very simple, yet effective, strategy is to exclude pixels with SB 
above a fixed threshold. 

\begin{figure}
\includegraphics[width=0.5\textwidth]{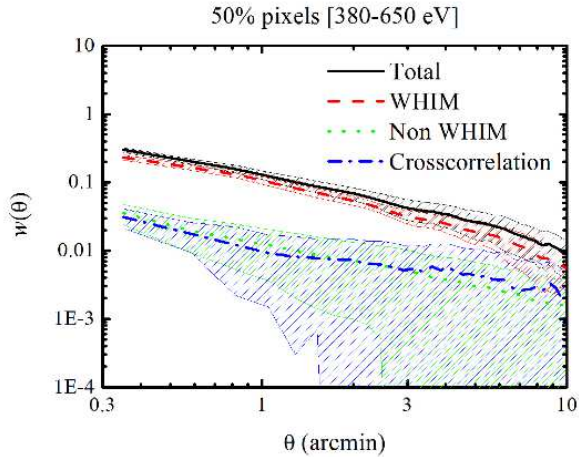}
\caption{Same as  Fig.~\ref{xenia_flux_cross} after removing 
the brightest $50\%$ pixels from the map.
\label{xenia_flux_cross_rem}}
\end{figure}

Fig.~\ref{xenia_flux_cross_rem} shows the result of having 
restricted the estimate of the ACF to $\sim50\%$ faintest 
pixels of the map. The result demonstrates the goodness of
our strategy: the ACF signal from the faint regions of the map is 
indeed dominated by the WHIM emission, as expected, and 
its ACF can be regarded as an indirect detection of the WHIM even in
absence of characteristic signatures in the correlation signal.
It is worth stressing that the method of discarding bright pixels is
a very rough one and can be easily improved, for example 
by targeting the pixels associated to known X-ray sources.
Yet, given the current uncertainties in the WHIM model and 
for the purpose of obtaining an order of magnitude estimate  
on the possibility of  extracting the WHIM correlation signal, 
we do not try to improve the ``cleaning'' procedure and focus instead 
on the problem of dealing with observational uncertainties.

\subsection{Realistic Case}
\label{acf-xenia}

We now repeat the analysis of \S~\ref{acf-ideal} using more realistic mock maps
which account for instrumental effects (exposure time and collecting area)
and for the presence of quasi-isotropic fore/backgrounds (Galactic foreground and
unresolved AGN).

\subsubsection{Discrete photon counts}
\label{acf-photon-counts}

In a realistic observational setup with finite exposure time and 
collecting area, one deals with photon counts. In order to transform 
the maps of \S~\ref{acf-ideal} into photon counts maps, we multiply 
the flux by the exposure time and the effective area of the detector. 
We consider observations with $t_{exp}$ in the range [0.1-1]~Ms and 
a detector of effective area $A=10^3$~cm$^2$, to mimic a long/very 
long exposure with the HARI ccd.

To generate the discrete photon maps we adopt a four-step procedure. 
As a first step we compute the total (integer) number of photons in 
the map as $N_{tot}=int\left(\sum_{ij}{f_{ij}A\,t_{exp}}\right)$, 
where $f_{ij}$ is the expected flux in the pixel with coordinates 
($i,j$) and the sum runs over all the pixels. In the second step we 
assign each pixel the corresponding (integer) number of photons 
$N_{ij}=int\left(f_{ij}A\,t_{exp}\right)$ and we keep track of the 
difference $P_{ij}=f_{ij}A\,t_{exp}-N_{ij} \in [0,1)$. In the 
third step, we assign the remaining $N_{tot}-\sum_{ij}N_{ij}$ 
photons through Monte Carlo rejection procedure that randomly 
assigns  photons to pixels with probability $P(i,j)$. Finally, we 
set the ratio of WHIM/{\it non}-WHIM photons equal to that of the 
WHIM/{\it non}-WHIM fluxes.

We did not add Poisson noise to the photon count maps since, for long 
exposure times, the error budget is dominated by field-to-field 
variance rather than shot noise. To justify this hypothesis we have 
evaluated the expected Poisson error using the following theoretical 
expression \citep{Cabre09}: 
\begin{equation}
\Delta\xi(r)=\sqrt{\frac{(N_R/N_D)^2+4(N_R/N_D)}{RR(r)}}, 
\label{eq:lserror}
\end{equation}
where $N_R$ and $N_D$ represent the number of emitters in the real 
and random catalog. In our case we have used  $N_R=N_D$. 

Here we focus on the WHIM signal in the maps from which we have 
removed the $50\%$ brightest pixels. In this case we find that 
field-to-field variance dominates the error budget for 
$t_{exp}\sim1$~Ms. As a result, the ACFs and its errors computed 
from the photon maps are identical to those computed from the flux 
maps, in the ideal case (e.g. Fig.~\ref{xenia_flux_cross}). However, 
when $t_{exp} \sim 100$~ks the Poisson noise contribution is no more 
negligible and accounts for $\sim 50\%$ of the error budget. To 
account for this effect we have increased the error bars on the 
corresponding maps by a factor $\sqrt{2}$. Instead, when we consider 
the {\it Total gas} signal in the original maps (i.e. with no bright 
pixel removal) the Poisson noise is largely subdominant, even for 
exposure times as short as 10 ks.

The same results holds true also when we remove the $50\%$ brightest 
pixels from the maps. It is not surprising that the ACFs obtained 
in this case give the same results as in the ideal map case 
(Fig.~\ref{xenia_flux_cross_rem}).

\subsubsection{Galactic Foreground}
\label{foreground}

Realistic DXB maps must account for the local contributions to the 
signal, i.e. for the Galactic foreground, which we model with two 
components: the Local Bubble and the Galactic Halo.
The Local Bubble emission, generated by the hot gas 
within $\sim200$~pc, is conveniently modeled as an unabsorbed thermal 
component with $\textrm{T}\sim10^6$~K and solar metal abundances 
\citep{Snowden97,McCammon02,Galeazzi07, Gupta09b}. In our maps
it is included using the APEC model in the XSPEC package. The 
Galactic Halo, generated by thermal emission of gas within our 
Galaxy, is well described by an absorbed thermal component with 
$\textrm{T}\sim2\times10^6$~K \citep{Snowden97,McCammon02,Galeazzi07}. 
It is added to our maps by combining an APEC and a WABS model in XSPEC. 
Both models use  the parameters listed in Tab.~3 of \citet{McCammon02}. 
The resulting SB of the Galactic foreground in the 
$380-650$~eV band is listed in Tab.~\ref{table_components}. 
This galactic foreground model is effective since it matches the
characteristics of the observed flux attributed to Galactic sources. 
Yet it is not accurate since it ignores the contribution from the 
Solar Wind Charge Exchange (SWCX) which has been shown to outshine 
that of  the Local Bubble \citep*{Koutroumpa06, Koutroumpa07, 
Henley08, Gupta09b, Koutroumpa09, Henley10}. The variable nature 
of the SWCX and the lack of strong observational constraints makes 
it difficult to model its contribution to the DXB. Fortunately, the 
very local origin of the  SWCX excludes the presence of small 
(i.e $\theta<10$') angular fluctuations so that its contribution to 
the Galactic foreground can be effectively modeled with an isotropic 
component which, following \citet{Gupta09b}, we obtain by boosting up 
the contribution of  the Local Bubble and Galactic Halo. 

Since angular variations of the Galactic foreground are expected 
to be significant only on angular scales much larger that those relevant for 
our correlation study, we model it as a purely isotropic component that 
contributes a $N_F$ photons in each pixel of the maps. For 
$t_{exp}=1$~Ms, $N_F=51.76$. Therefore, we obtain the 
number of Galactic foreground photons in the maps 
by Monte Carlo sampling a Poisson distribution with mean $N_F$.

The addition of the Galactic foreground significantly reduces the possibility
of detecting the WHIM signature in the DXB in two ways.
First, it decreases the amplitude of the ACF signal. According to
Eq.~\ref{add_acf} the expected decrement is $\sim50\%$ 
in the case of $t_{exp}=1$~Ms.
The effect on the signal-to-noise error, however is less dramatic since the 
presence of  a bright isotropic foreground 
greatly reduces the field-to-field scatter which dominates the noise. 
In fact, the overall reduction of the statistical 
significance of the correlation signal is just $\sim20\%$. 
Second, and more serious, is the impact of the Galactic foreground on 
our capability of extracting the WHIM signal from the maps. 
The addition of a bright foreground outshines the correlation signal
in the faint pixels. As a consequence, 
removing the brightest $50\%$ pixels results in 
shot-noise dominated maps with
zero ACF  at all angles, as we have verified.

\begin{figure}
\includegraphics[width=0.5\textwidth]{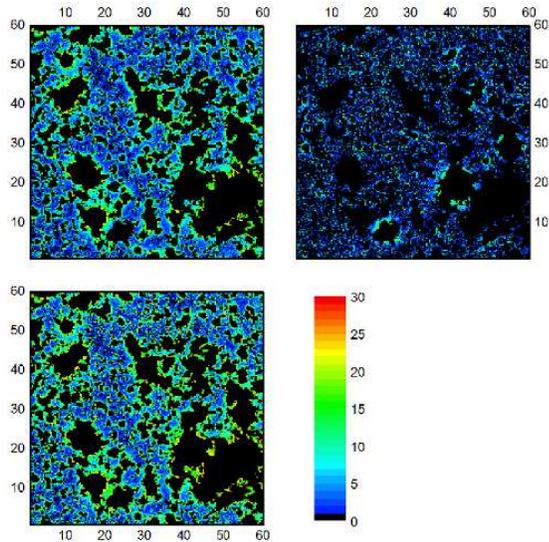}
\caption{Residual maps of WHIM (\emph{top-left}), {\it non}-WHIM 
(\emph{top-right}), and {\it Total gas} (\emph{bottom-left}) in the 
$380-650$~eV energy band for a f.o.v. of $1\times1$~deg$^2$, 
effective area of 1000~cm$^2$, and $t_{exp}=1$~Ms. The color scale 
is in photon-number units. 
\label{maps_bright}}
\end{figure}

A simple solution to this problem is that of removing a model  
Galactic foreground before analyzing the ACF of the maps. Foreground 
removal would be then performed by simply subtracting the 
Galactic foreground model (a random counts map
characterized by its mean intensity) to the photon count map.
Since we assume that random counts
are Poisson distributed, the subtraction is performed as follows.
First we Monte Carlo generate the number of random counts in 
the pixel, $F_{ij}$. Then we compute the residual map
$R_{ij}=T_{ij}-F_{ij}$ with the constraint that $R_{ij}=0$ if $T_{ij}<F_{ij}$. 
Since this procedure removes less photons 
from the total map than those in the foreground  map $F_{ij}$,
we randomly remove the remaining photons from 
the pixels with $R_{ij}>0$. Finally, we remove the 50\% brightest 
pixels from this residual map. The result is displayed in 
Fig.~\ref{maps_bright}. The panels show the same f.o.v. as 
Fig.~\ref{maps}. The extended, black areas correspond to the regions 
in which the 50\% brightest pixels have been removed from the map.
Isolated black pixels in the low-count regions are those with 
$R_{ij}=0$.

\begin{figure}
\includegraphics[width=0.5\textwidth]{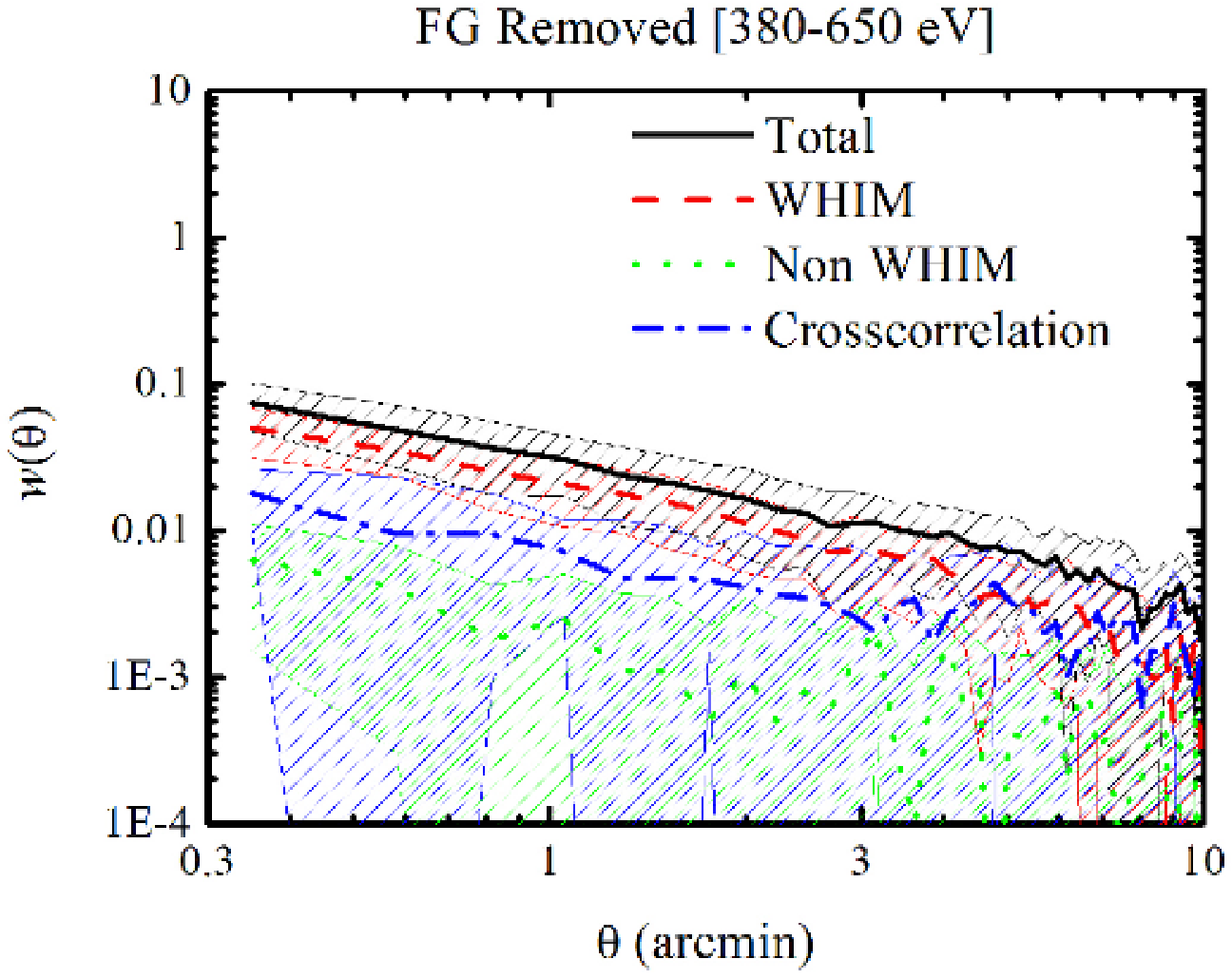}
\includegraphics[width=0.5\textwidth]{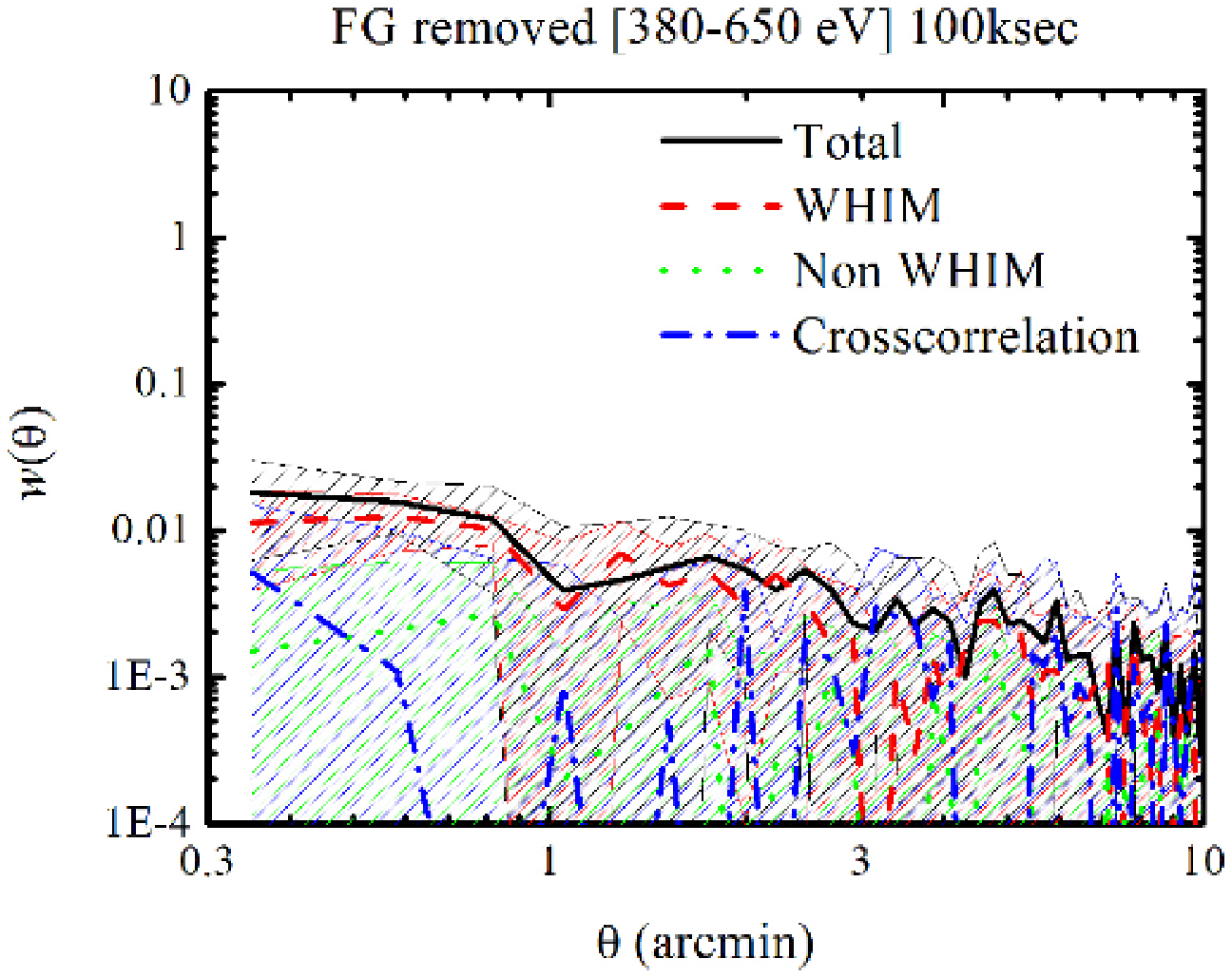}
\caption{Comparison between the ACF of total map (\emph{solid black}), 
WHIM component (\emph{dashed red}), {\it non}-WHIM component 
(\emph{dotted green}), and the cross-correlation (\emph{dot-dashed 
blue}). All ACFs are normalized to the total flux after removing the 
brightest $50\%$ of the pixels. The plots refer to the case of a 
$t_{exp}=1$~Ms observation (\emph{top}) and a $t_{exp}=100$~ks 
observation 
(\emph{bottom}). 
\label{countclean_fgrem}}
\end{figure}

This  procedure to remove the foreground is very simplistic but, as 
shown in  the top panel of Fig.~\ref{countclean_fgrem} it is good 
enough to extract the WHIM autocorrelation signal from a 
$t_{exp}=1$~Ms observation, since the ACF below $\theta\sim3$' is 
significantly above zero and dominated by the WHIM component (red 
dashed curve). 

As in the previous section, we do not add Poisson noise to the WHIM 
and {\it non}-WHIM counts. This choice is justified because, in 
addition to the fact that cosmic variance dominates over the Poisson 
noise of the signal for f $t_{exp}=1$~Ms, as discussed previously, the 
Poisson noise is now dominated by the Foreground, not by the signal, 
which is included in our maps.

Given the uncertainties in the WHIM model adopted, these results are 
to be regarded as indicative. Yet, they serve to illustrate an 
important point: the capability of extracting the WHIM autocorrelation 
signal from the DXB maps is mainly hampered by the presence of a 
bright Galactic foreground. With this respect, our problem is 
analogous to that of extracting a weak signal in presence of a strong 
random noise. In this case, the signal-to-noise can be improved in 
two ways. 

Since, in the large noise limit, signal-to-noise ratio scales as 
$t_{exp}$, the simplest option is to increase the exposure time. The 
effect of decreasing $t_{exp}$ by a factor 10 is illustrated in the 
bottom panel of Fig.~\ref{countclean_fgrem}. In this case the ACF 
signal has been washed out everywhere but at sub-arcmin angular 
separations. Varying the exposure time should not affect the ACF 
signal but only its uncertainties, if they are dominated by Poisson 
noise. However, the two panels of Fig.~\ref{countclean_fgrem} show 
that this is not the case. We have verified that this apparent 
inconsistency reflects a systematic bias introduced by our procedure 
to remove the foreground from a map in which the signal is dominated 
by the foreground itself. The shorter the exposure time, the less 
efficient the foreground removal, the more Poisson noise dominates 
the residual map. The net results is to wash out the ACF signal in 
the maps with shorter exposure times.

To assess the accuracy of our simple strategy to remove the foreground 
we have considered an additional procedure in which we remove a 
constant signal with amplitude equal to the mean foreground, but with 
no Poisson noise included, with the caution of avoiding negative 
photon counts. The results are basically the same as with the 
original procedure.

Rather then increasing the exposure time, a better alternative is to 
improve the map cleaning efficiency. First of all, one can perform a 
more effective exclusion of the {\it non}-WHIM signal by excluding 
(or filtering out) the signal in correspondence  of known X-ray 
sources. Second of all, a better extraction of the WHIM signal can 
be obtained by convolving the map with Wiener-like filters 
\citep*{Wiener49, Press92}. Of course Wiener filtering relies on 
{\it a-priori} modeling of the WHIM ACF and Galactic foreground 
noise. However, possible inaccurate values of the Wiener Filter 
resulting from model uncertainties would introduce second order 
errors only \citep{Rybiky92}.

Our analysis, therefore, clearly shows that long exposures are 
required to extract the WHIM autocorrelation signal from the ACF of 
the DXB. However, our considerations suggest that the value of 
$t_{exp}=1$~Ms is probably too generous and that a significant 
detection might be possible with shorter exposures. This conclusion 
is corroborated by the recent results of \citet{Galeazzi09}, who 
studied the ACF of the DXB in a set of {\it  XMM}-fields with 
$t_{exp}$ ranging from $50$~ks to $600$~ks. After removing the 
contribution from known sources and making no attempt to subtract the 
Galactic foreground, they detect a positive autocorrelation signal 
in all fields (Fig.~2 of \citealt{Galeazzi09}) that they interpret 
as the signature of the WHIM.

Finally, we comment on the error contribution from the expected 
instrumental noise. The baseline of \Xenia CCD detector is similar to 
that of {\it Suzaku} and relies on being placed on a low-Earth orbit. 
The rational behind this choice is to guarantee a more stable 
background than that in highly elliptical orbits where {\it Chandra} 
and {\it XMM-Newton} have been placed. The {\it Suzaku} instrumental 
background is $\sim2\times10^{-5}$~counts~kev$^{-1}$~sec$^{-1}$~mm$^{-2}$ 
for XIS-FI and $\sim5\times10^{-5}$ ~counts~kev$^{-1}$~sec$^{-1}$~mm$^{-2}$ 
for XIS-BI, i.e. 4 times and 9 times smaller than the cosmic X-ray 
background \citep{Tawa08}. By comparison instrumental backgrounds for 
XMM-MOS and for XMM-PN are 2 times and 5 times smaller than the cosmic 
X-ray background. 

\subsubsection{Resolved and unresolved AGN}
\label{agn}

We also include the contribution of resolved and unresolved AGN in 
the mock maps. 

Resolved AGN can be easily eliminated from the maps by excluding the 
contaminated pixels. However, given the limited number of expected 
objects, the effect is rather small. For $t_{exp}=1$~Ms, \Xenia is 
expected to resolve $\sim2000$ AGN per square degree above the 
detection threshold $\textrm{S}_{Xenia}$ of $10^{-16}$~erg~cm$^{-2}$
~s$^{-1}$ in the $0.5-2$~keV band \citep{Piro09}. Under the somewhat 
simplistic assumption that each resolved AGN occupies just one pixel, 
the fraction of excluded pixels is just $3\%$ of the total, much 
smaller than the $50\%$ fraction of bright pixels that were excluded 
to remove the {\it non}-WHIM contribution. It is no surprise that 
removing this extra $3\%$ of pixels has no appreciable impact on our 
result. It is also clear that a more sophisticated model, in which 
one account for the instrumental PSF, would not affect our results 
either. In fact the situation is even more favorable, since most of 
the AGN will be associated to massive structures contributing to the 
{\it non}-WHIM signal corresponding to the already removed $50\%$ 
brightest pixels.

To estimate the effect of unresolved object, we have included their 
contribution to the DXB by generating a population of unresolved 
sources matching the number counts and redshift distribution of faint 
AGN observed by {\it Chandra} and using N-body simulation to model 
their spatial correlation properties. More specifically, we generate 
a mock population of AGN as follows.

First we assume that mock AGN have the following number counts:
\begin{equation}
N(>S) \left\{ \begin{array}{ll}
S^{-1.5}, & \propto S_{Lim}<S<S_{Chandra}\nonumber \\
S^{-0.6}, & \propto S_{Chandra}<S<S_{Xenia} 
\end{array} \right. ,
\label{eq:lognlogs}
\end{equation}
where $S_{Chandra}=3\times10^{-17}$~erg~cm$^{-2}$~s$^{-1}$ is 
similar to the 
detection limit in the Chandra Deep Fields and 
$S_{Lim}=6.5\times10^{-18}$~erg~cm$^{-2}$~s$^{-1}$ is a lower limit so 
that the function does not diverge at infinity. In practice we require that mock 
AGN counts match the {\it Chandra} ones above $S_{Lim}$
\citep*{Moretti03}
and assume that they follow the usual Euclidean slope below 
$S_{Lim}$. The reason of assuming a Euclidean slope rather than extrapolating
the one of the resolved AGN is to maximize the contribution 
of unresolved objects to the DXB. 
To assign flux to the mock unresolved AGN we sampled the 
probability distribution obtained from Eq.~\ref{eq:lognlogs}
between  $S_{Xenia}$
to $S_{Lim}$. The total number of object is determined by requiring 
that unresolved AGN contribute to $\sim 12\%$ of the Cosmic X-Ray Background
(assuming a Cosmic X-Ray Background flux of 
$7.5\times10^{-13}$~erg~cm$^{-2}$~s$^{-1}$~deg$^{-2}$ \citealt{Moretti03}). 
In Tab.~\ref{table_components} we report the flux of the unresolved 
AGN in the $0.38-0.65$~keV energy band assuming a spectral slope $\Gamma=1.4$. 
We end up by generating
 $\sim5\times10^4$ objects per square degree 
with $S_{Lim}\le S < S_{Xenia}$.
 
To determine the redshift distribution $\frac{dN}{dz}$
of these objects 
we integrate the AGN luminosity function of \citet{Hopkins06}
specified in three different redshift bins, out to $z=3$. 

Finally, we determine the spatial location of the mock AGN from the 
outputs of a collisionless N-body experiment under the assumption 
that AGN are located at the centre of dark matter halos. In practice, 
we extract 10 independent light-cones with f.o.v. of 
$1^\circ\times1^\circ$ from the publicly available GIF simulations 
\citep*{Jenkins01}. Light cones are obtained by stacking all 
available outputs of the 479~Mpc~h$^{-1}$ computational cube out to 
$z=3$. Within each cone we adopt a Monte Carlo rejection procedure 
to select a subsample of $5\times10^4$ dark matter halos with mass 
$>1.4\times10^{12}$~M$_\odot$~h$^{-1}$, enforcing the required 
$\frac{dN}{dz}$. X-ray fluxes are randomly assigned according to 
Eq.~\ref{eq:lognlogs} and the angular position of the AGN within each 
light-cone is obtained by projecting the spatial position of each 
selected halo on the $1^\circ\times1^\circ$ f.o.v.

By construction, mock AGN have the same autocorrelation function of 
the dark matter halos, but no cross-correlation with the diffuse gas, 
since gas particles and dark matter halos have been taken from 
different simulations.

This inconsistency, however, is hardly an issue since, as we have 
verified, the angular autocorrelation function of the unresolved AGN 
is consistent with zero within the errors. We conclude that refining 
the AGN model to account for the spatial correlation with the gas 
distribution would not change our results. Unresolved AGN fainter than 
S$_{Lim}$ provide no significant correlation signal and can be 
regarded as an additional, weak isotropic signal to be added to the 
Galactic Foreground \citep{Plionis08}.

Another potential class of sources that may contribute to the DXB 
is that of normal galaxies, thanks to their X-ray coronae and X-ray 
binary population. However, at the \Xenia detection threshold for 
1~Ms exposure, $10^{-16}$~erg~cm$^{-2}$~s$^{-1}$, their contribution 
is largely dominated by that of the AGN, as shown e.g. by 
\citet{Hickox06}. To independently check the validity  of this 
conclusion we have extrapolated the Log~{\it N}/Log~{\it S} of normal 
galaxies computed by \citet*{Georgakakis06} (Fig. 3 of their paper) 
down to $6.5\times10^{-18}-1\times10^{-16}$~ergs~s$^{-1}$~cm$^{-2}$ 
and computed their contribution to the diffuse background of \Xenia. 
In this exercise we had to transform the original flux in the 
$[0.5-2]$~keV band into the X-ray band considered in this paper, 
$[0.380-0.650]$~keV. For this purpose we have assumed that the 
typical $0.380-0.650$/$0.5-2$~keV flux ratio for a normal galaxy 
is identical to that of the Milky Way. A choice that probably 
provides an upper limit since the the many Galactic lines that 
contribute to the flux in the $[0.380-0.650]$~keV are redshifted out 
of this range in the spectrum of a distant object. We find that 
the contribution of unresolved galaxies is 
$2.78\times10^{-4}$~phot~s$^{-1}$~cm$^{-2}$~sr$^{-2}$, i.e. 
a factor of $\sim 2$ smaller than that of the AGN. Had we assumed a 
more conservative, single power law  $N(>S)\propto S^{-0.6}$ for 
the unresolved AGN, as e.g. in \citet{Hickox06}, the relative 
contribution of normal galaxies would have been larger.

At lower fluxes the contribution of normal galaxies is expected to 
dominate over the AGN, but not over that of the WHIM, as discussed 
in \citet{Hickox07} and detailed in Table~\ref{table_components}.

\section{Tracing the spatial distribution of the WHIM}
\label{3DACF}

In this section we explore the possibility of performing
2D spectroscopy to characterize
the 3D distribution of the WHIM. 
For this purpose we take the CRIS instrument as reference 
and use our mock 2D spectra to assess how well one can trace
the 3D distribution of the WHIM using its 
characteristic O\elem{VII} and O\elem{VIII} emission lines.
The angular and energy resolution 
of the instrument limits the ability of tracing the 
angular and redshift distribution of the WHIM, respectively.
The exposure time determines the sampling density of the line-emitting gas.
Here we consider the case of $t_{exp}=1$~Ms.

The number of WHIM detections
expected with an instrument like CRIS has been estimated by \citet{Takei10}.
Their analysis shows that the minimum line surface brightness required
for a  $5\sigma$ detection is  0.07\phot and that the expected 
number of simultaneous O\elem{VII} and O\elem{VIII} detections 
is 639 per square degree, 426 of which are contributed by the WHIM
(see Tab. 2 of \citealt{Takei10}).
In fact, the actual number of detections is 
expected to be significantly larger, since these estimates were obtained 
considering gas with $z<0.5$, whereas the WHIM mass fraction is 
still large up to $z \sim 1$. 

These detection estimates, however, have been obtained for an angular 
resolution ($2.6'\times2.6'$) larger than that of our mock 2D spectra 
($1.3'\times1.3'$).
As a consequence, we should revise detection estimates to account for 
the larger number of angular resolution elements available and for the 
larger threshold (0.22\phot) required for a  $5\sigma$ detection.
However, our goal is to detect line emitters that were
obtained from a grouping procedure. It turns out that typical angular size
of the emitters is closer to $2.6'\times2.6'$ 
than to $1.3'\times1.3'$ and we decide to keep the same 
threshold as \citet{Takei10} to select detectable emitters.
The fact that the number of detectable emitters is close to the 
estimates of \citet{Takei10} confirms the goodness of this choice.

To characterize the spatial distribution of the line emitters we use the 
spatial two-point correlation function $\xi(r)$, which we compute 
using the same estimator used for the angular correlation function
(see Section \ref{AcF-test}), generalized to the three dimensional case. 
The analysis is performed both in real and redshift space, i.e. by 
considering both comoving distances and ``observed'' redshifts.
Finally, as already pointed out, since for our emitters the presence of 
a detectable O\elem{VIII} line often guarantees the simultaneous detection
of the O\elem{VII} line, in our analysis we will only consider
O\elem{VIII} emitters.

\subsection{Redshift distribution of line emitters}
\label{cum_emitters} 

Before studying the spatial correlation properties of the O\elem{VIII} emitters, 
let us analyze their redshift distribution out to the  comoving distance 
$d_c \simeq$ 1300~Mpc~h$^{-1}$, corresponding to the maximum
redshift $z=0.5$. In this redshift range the WHIM mass fraction
is expected to increase by only $\sim10\%$, whereas the mass fraction 
of the hot gas ($T>10^7$~K) should increase by $\sim50\%$ \citep{CenOst06}. 
If O\elem{VIII} lines were mostly produced 
by hot gas in virialized structures we  should observe a significant evolution
in the number density of  O\elem{VIII} emitters. On the contrary, if emission were dominated 
by the WHIM, the emitter number density should remain constant.
Here we briefly remind the reader of the definition adopted in Section 
\ref{Model}. The {\it Total Gas} consists of all gas with temperature 
above $10^5$~K. The {\it Bright}-WHIM represents where the WHIM 
contributes to the X-ray emission is detectable (regardless of the 
non-WHIM). The WHIM-{\it dominated} gas consists of gas regions in which 
the WHIM contribution to the line emission is larger than that of 
all other phases.

\begin{figure}
\includegraphics[width=0.5\textwidth]{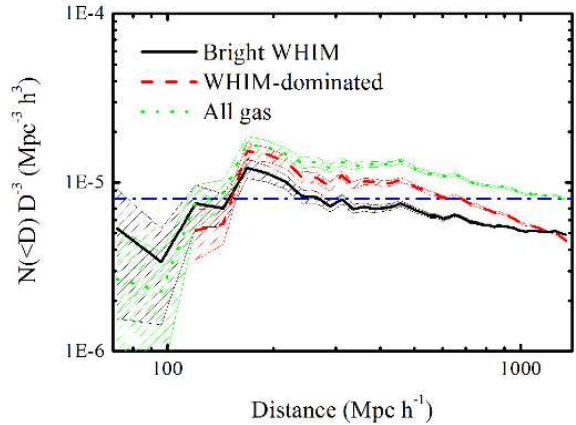}
\caption{Cumulative distribution of O\elem{VIII} emitters normalized to 
that of a population of objects with constant number density
 $\propto d_c^3$.  {\it Bright} WHIM (\emph{solid, black}), WHIM-{\it dominated} 
(\emph{red, dashed}) and  {\it Total gas} (\emph{green, dotted}) emitters. 
The \emph{blue, dash-dotted} line is a reference distribution for a 
volume limited sample ($N\left(<d_c\right)\propto d_c^3$).
\label{cum_real}}
\end{figure}

Fig.~\ref{cum_real} shows the cumulative density distribution of the 
O\elem{VIII}-line regions in the simulated light-cone as a function of 
the comoving distance $d_c$.
Once we ignore the local region 
($d_c<$ 300~Mpc~h$^{-1}$) dominated by a few structures
and characterized by a large variance, 
we see that the comoving number density of all  O\elem{VIII} emitters 
(green-dotted curve) is not too different from that of a volume limited sample,
$N\left(<d_c\right)\propto d_c^3$ (blue dash-dotted). Deviations from 
this law exist for $d_c>$ 500~Mpc~h$^{-1}$.  
However, the fact that the slope of the {\it Bright}-WHIM
curve (black, solid) is similar to that of the {\it Total Gas}
suggests that they are not due to evolutionary effects.
In fact, the similar shapes of the curves indicates that
in both cases the emission is dominated by diffuse material, whose mass fraction
hardly changes in this redshift interval, rather than by virialized gas.
This result simply confirms that, due to 
the unfavorable ionization balance, the bulk of the O\elem{VIII}-line 
emitters does not trace the hot ($T>10^7$~K) gas associated to 
galaxy clusters. Instead, they  trace 
the gas located in the outskirts of virialized objects.
Deviations from the $N\left(<d_c\right)\propto d_c^3$ curve simply derives from
the fact that the volume of the resolution elements increases with the distance.
This means that two nearby line-emitting regions along the 
same line of sight 
would be counted as two different emitters when they are close to the observer but would 
count as a single object at large distances. 

The very fact that the cumulative number density of the WHIM-{\it dominated} 
(red, dashed)
emitters decreases faster with the distance than that of the {\it Bright}-WHIM emitters
corroborates this hypothesis. To understand this point, one has to 
keep in mind that the {\it Bright}-WHIM gas is found in the same regions, the outskirts
of virialized objects, as the WHIM-{\it dominated}. The main difference is that the latter is, on 
average, considerably fainter, i.e. its lines are closer to the detection threshold.
Therefore an extended region that looks like a single spot, if regarded as  {\it Bright}-WHIM,
would have a more patchy appearance when regarded as WHIM-{\it dominated} gas.
The grouping procedure might then extract several WHIM-{\it dominated} emitters 
from a region in which only a single  {\it Bright}-WHIM emitter is found.
The magnitude of the oversampling decreases with the redshift 
as demonstrated by the fact that in Fig.~\ref{cum_real} the total 
number of {\it Bright}-WHIM emitters matches that of the 
WHIM-{\it dominated} emitters at $d_c>$ 1000~Mpc~h$^{-1}$.
 
To summarize, the cumulative redshift distribution of the O\elem{VIII} emitters 
suggests little or no evolution in the range $z\in[0,0.5]$. 
Deviations from a population of objects with constant number density 
and discrepancies between WHIM-{\it dominated} and
{\it Bright}-WHIM emitters can be understood in terms of geometry effects. 
As fully explained in the following section, the errors shown in 
Fig.~\ref{cum_real} are not estimated as scatter between different 
samples since we have only one catalog, instead we assumed a Poissonian 
distribution of emitters, and the errors are simply the square root 
of emitters' number.

\subsection{Two-point correlation function in real space}
\label{3d:corr}

\begin{figure}
\includegraphics[width=0.5\textwidth]{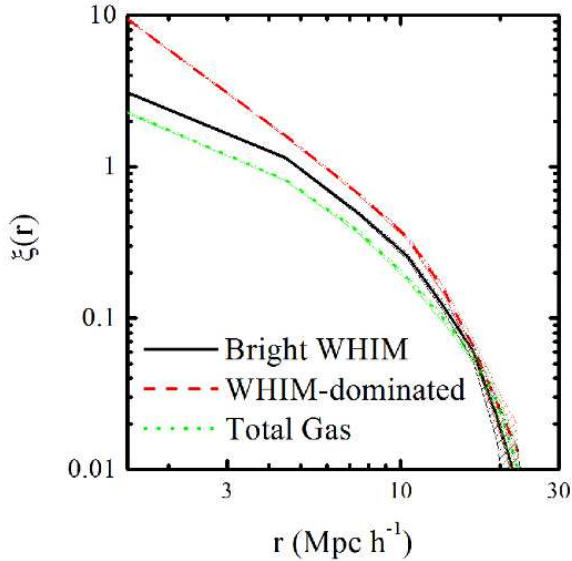}
\caption{Real-space correlation function for {\it bright}-WHIM 
(\emph{solid black}),  WHIM-{\it dominated} (\emph{dashed red}), and 
{\it Total gas} (\emph{dotted green}) O\elem{VIII} emitters. Shaded areas 
represent Poissonian uncertainties.
\label{xireal}}
\end{figure}

Fig.~\ref{xireal} shows the spatial two-point correlation function, 
$\xi(r)$, of the detectable O\elem{VIII}-line emitters as a 
function of the comoving pair separation. The correlation function 
is measured in real space, ignoring peculiar velocities. The curves 
refer to the three families of emitters: {\it Bright} WHIM (black, solid),
WHIM-{\it dominated}  (red, dashed) and {\it Total gas} (green, dotted).
The correlation functions are computed in comoving bins of 
3~Mpc~h$^{-1}$  to account for the instrument energy resolution. 
Ideally, we would like to estimate errors from the scatter among the mocks,
as in \S~\ref{2DACF}. However, in this case we only have one mock catalog at our disposal.
Fortunately, given the large volume covered by  our mock 2D spectra and the limited number of 
detectable  emitters, we expect that the errors in the $\xi(r)$
are dominated by sparse sampling rather than by field-to-field variance.
To model shot noise we use Eq.~\ref{eq:lserror}, as we did in 
Section~\ref{acf-photon-counts}, this time adopting $N_R/N_D=3$.
In this work we rely on Eq.~\ref{eq:lserror} rather than computing errors using 
 more sophisticated techniques (bootstrap resampling,  jackknife) since  the 
 main goal here is to obtain an order-of-magnitude estimate for the errors
 to assess our future capability of characterizing the 
 correlation properties of the WHIM. In practice we will only trust 
 differences that are significantly larger than the 
 1-$\sigma$ width of the error-bands in  Fig.~\ref{xireal}
  
The main result of our analysis, shown in
Fig.~\ref{xireal}, is that O\elem{VIII} emitters, whether or not contributed by the WHIM,
are expected to be spatially correlated and  their two point correlation function
will be detected with high statistical significance
by next generation experiments like {\it Xenia}. 
The large signal-to-noise ratio in the expected $\xi(r)$ demonstrates the robustness 
 of this prediction which cannot be too much affected by uncertainties in the WHIM model
 and approximations in our error estimates.
 
All two-point correlation functions in Fig.~\ref{xireal} are well approximated 
by a power law with slope $\gamma=-1.7\pm0.1$ 
in the range $r=[4.5,12]$~Mpc~h$^{-1}$, steepening at 
larger separations. Such slope, determined via $\chi^2$ minimization, is common to
many cosmic structures, from galaxies to clusters and simply shows that both 
the O\elem{VIII} emitters and virialized objects trace the underlying distribution of
Dark Matter, as expected in a standard CDM scenario.

In particular, the correlation length of all detectable emitters (the {\it Total gas})
$r_0=4.0\pm0.1$~Mpc~h$^{-1}$ is consistent with that of 
optically selected galaxies as measured in several redshift 
catalogs \citep{Zehavi02,Hawkins03}. Such coincidence does not 
necessarily indicate a 1:1 relation between galaxies and  O\elem{VIII} emitters.
Instead, it shows that both types of objects trace the same 
dark matter structures at separations
$>3$~Mpc~h$^{-1}$, significantly larger than the virial radius 
of galaxy groups and clusters.

One interesting feature of Fig.~\ref{xireal} is that 
the correlation of the emitters contributed by the WHIM
 ({\it Bright}-WHIM and WHIM-{\it dominated}) is stronger than that of the 
{\it Total gas}, whose line emission is contributed by gas in groups and clusters.
This result is counter-intuitive since hydro-simulations show that 
the WHIM preferentially traces the filamentary structure of the cosmic web whose 
spatial correlation is weaker than that of  virialized objects. 
This results can be understood as the combination of two effects:
a selection effect, derived from having imposed a given SB threshold,
and the peak-biasing phenomenon \citep{Kaiser84}, which results 
from sampling the peaks of the underlying mass  density field.
In the case we are considering, the O\elem{VIII}-line associated to 
the WHIM in filamentary structures is too weak to be detected. 
Only the WHIM in the outskirts of large, virialized structures
is O\elem{VIII}-bright enough to be detected. In other words: 
with our selection we only sample {\it Bright}-WHIM and WHIM-{\it dominated}
emitters associated to galaxy clusters, whereas {\it Total gas} emitters
are preferentially associated to the external regions of smaller, galaxy group-size 
structures.
Galaxy clusters are associated to the highest peaks of the mass density fields,
higher than those associated to galaxy groups. The higher the peaks, the larger their 
two-point correlation function \citep{Kaiser84}, which explains the stronger correlation 
of WHIM-contributed O\elem{VIII} emitters.

We stress that the explanation above, though generally valid, has to be regarded 
as semi-quantitative since O\elem{VIII} emitters are not directly associated to virialized
structures, but to their outskirts, i.e. regions beyond their virial radius. For this reason 
it is not possible to interpret our result in terms of a halo biasing model, 
and infer the typical properties of dark matter halos from the correlation function of the 
 O\elem{VIII} emitters.
 
In Fig.~\ref{xireal} we also notice that the  WHIM-{\it dominated} emitters are more 
correlated than the {\it Bright}-WHIM ones. The reason for this difference  is more 
subtle. We have seen that both types of  emitters 
 sample the same regions (the outskirts of galaxy clusters) and so they should 
 have the same correlation function. However, as we have discussed in 
Section~\ref{cum_emitters},  the WHIM-{\it dominated} emitters oversample these
regions due to their patchy appearance. The result of oversampling the very same 
peaks of the density fields is to increase the correlation of the WHIM-{\it dominated}
with respect to that of the {\it Bright}-WHIM emitters.

The small scale  ($\le 3$~Mpc~h$^{-1}$) flattening of the correlation function
is an artifact. It reflects the fact  that the energy resolution of the 
instrument effectively acts as an anisotropic smoothing filter with radial 
scale of $3$~Mpc~h$^{-1}$ which erases correlation on the corresponding scales.
Once again, the different behavior of the WHIM-{\it dominated} emitters reflects 
their oversampling of the density peaks, which is particularly effective at small
separations.

\begin{figure}
\includegraphics[width=0.5\textwidth]{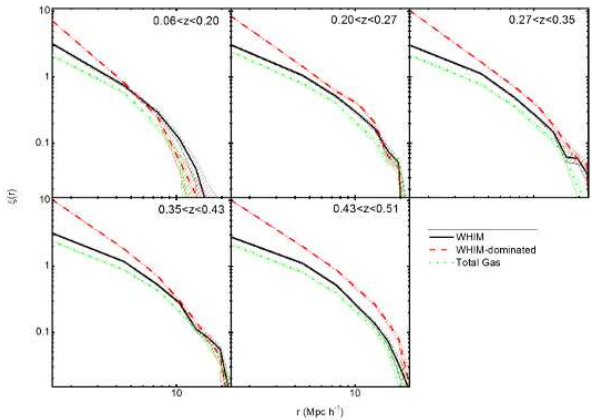}
\caption{Real-space correlation function for {\it bright}-WHIM, 
(\emph{solid black}),  WHIM-{\it dominated} (\emph{dashed red}), and 
{\it Total gas} (\emph{dotted green}) O\elem{VIII} emitters in six redshift 
slices. The redshift range is indicated in each panel.
\label{xirealdiff}}
\end{figure}

Fig.~\ref{xireal} shows the correlation function
measured in the full light-cone. 
Since the mass fractions of different gas phases evolve differently 
we should check whether their correlation functions evolve in a different way too.
To investigate this issue, we have measured the 
spatial correlation function of the O\elem{VIII} emitters 
slicing the interval $z=[0,0.5]$ 
into six bins. The results are shown in Fig.~\ref{xirealdiff}. 

There is little evolution in the correlation properties of the
emitters regions in the interval explored, with the exception of the 
nearest and outermost slices. At $z\sim0.5$ the WHIM-{\it dominated} 
regions appear to be more correlated on larger scales than at lower 
redshifts. The simple explanation is, again, related to selection effect and peak biasing.
At large redshift one detects stronger O\elem{VIII} emitters
associated to higher density peaks and the amplitude of the correlation function increases accordingly.
On the opposite, at $z\sim0$
all types of emitters are less correlated than at higher redshifts. The effect is particularly visible
at large separations but its statistical significance is not large. The point is that the volume 
associated to the first slice of the cone is rather small and  does not allow 
to fairly sample scales $\geq10$~Mpc~$h^{-1}$. Therefore, the error budget is 
likely to be dominated by cosmic variance  rather than shot-noise errors. 
Adding the cosmic variance contribution would increase the size of the 
error-bands in Fig.~\ref{xireal} and reduce the statistical significance of the effect.

\subsection{Correlation analysis in redshift space}
\label{redshift_space:corr}

From the analysis of the X-ray spectra one infers the 
3D distribution of the O\elem{VIII} 
emitters in redshift space, not in real space.
The presence of peculiar velocity spoils the 1:1 relation between redshifts and 
comoving distances and introduces systematic distortions in the 
spatial clustering of the emitters and in their correlation properties.
The net result is that the two-point correlation function measured in redshift space
is systematically different from the true one, measured in real space. The 
mismatch can be used to infer the properties of the peculiar velocity field, i.e.
to characterize the dynamical state of the emitters.

Peculiar motions can be roughly divided in two categories.
Incoherent motions, which typically dominate at small separations, and 
represent random velocities within virialized structures. They cause these 
structures to appear  elongated along the line of sight in a characteristic
type of distortion commonly known as ``Fingers of God''.
Coherent motions characterize large scale structures  growing in 
the linear or quasi-linear regime and typically represent infall motions 
into mass overdense regions (or outflow from underdense regions).
They apparently increase the density contrast associated to these regions along the 
line of sight.

For a collisionless fluid, like the  galaxies in the universe, measuring these motions allows 
to infer important properties of the underlying mass density field: the amplitude of virial motions 
allows to infer the typical mass of virialized structures in the sample, while the size of coherent motions
allows to constrain the growth rate of density fluctuations \citep*{Hamilton92, Guzzo08} 
from which one can measure the mass density parameter $\Omega_m$ \citep*{peacock01}
or test alternative gravity models \citep{Linder05}.
For the case of a collisional dissipative fluid, like the O\elem{VIII} emitting gas, 
one cannot derive these parameters from redshift distortions.
However, it is still possible to infer the dynamical 
status of the gas which still brings some important information on the typical environment in
which the WHIM is found. 
A strong Finger of God signature  would indicate that the gas is 
preferentially found within virialized environments, while a 
dominance of the compression effect by coherent motions would indicate 
that the WHIM is preferentially found in regions of moderate overdensity 
moving toward virialized regions or outflowing from low density environments.

\begin{figure}
\includegraphics[width=0.5\textwidth]{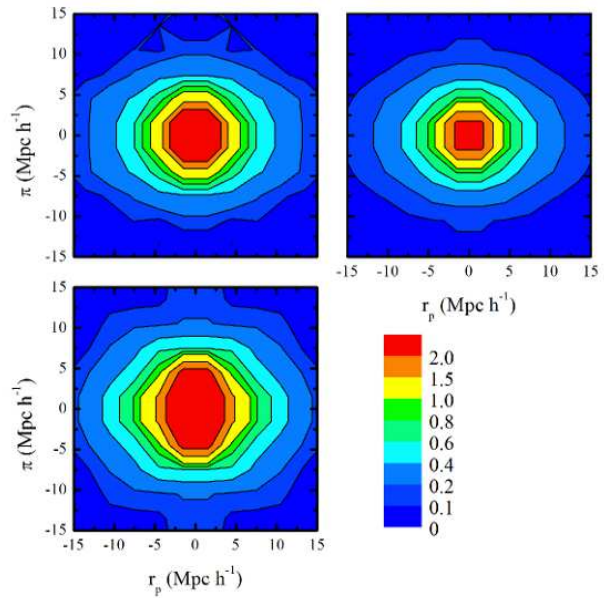}
\caption{Iso-correlation contours for the measured two-point 
correlation function $\xi$ in the plane $(r_p,\pi)$ in redshift-space. 
The three panels refer to the three types of emitters considered.
{\it Bright}-WHIM (\emph{top-left}). {\it Total gas} (\emph{top-right}).
WHIM-{\it dominated} (\emph{bottom-left}) 
The color scale quantifies the amplitude of the 
iso-correlation contours.
(\emph{bottom-right}).
\label{isoall}}
\end{figure}

Peculiar velocities cause distortions only along the line of 
sight, and not on the transverse direction. One can therefore 
quantify their amplitude by comparing the clustering of objects 
in the directions parallel and perpendicular to the line of sight.
For this purpose  it is convenient to compute the two point 
correlation function after decomposing pair separations $r$ 
into the line of sight component, $\pi$, and perpendicular to it, $r_p$. 
The resulting correlation function 
$\xi(r_p,\pi)$ is readily computed 
using the usual LS estimator. The three panels in Fig.~\ref{isoall} 
show the two-point correlation  function $\xi(r_p,\pi)$ 
of the three types of
 O\elem{VIII} emitters, projected onto the $r_p,\pi$ plane.
Different colors indicate different iso-correlation contours, 
whose amplitude  is indicated  in the color scale. 

With no peculiar velocities, clustering properties are expected to be
isotropic and iso-correlation contours would be circularly symmetric.
Deviations from the circular symmetry would then indicate 
the presence of peculiar-velocity induced anisotropies in the
spatial clustering of the emitters. The stronger the deviations, the 
larger the amplitude of peculiar velocities.
As anticipated these distortions can roughly be divided in two 
categories. Incoherent velocities associated to Fingers of God
type distortions cause the iso-correlation contours to be elongated along
the $\pi$ direction. Being associated to small-scale motions they constitute the 
dominant type of distortions at small values of $r_p$.
 Coherent motions associated 
to enhancements of density contrasts induce a compression of 
the iso-correlation contours along  $\pi$.
Since they are associated to large-scale motions, they 
dominate the distortion pattern at large values of  $r_p$.

The iso-correlation contours shown in Fig.~\ref{isoall}
display a non-negligible compression for $r_p\ge6$~Mpc~h$^{-1}$
whose magnitude does not seem to depend on the 
type of emitter.
This result suggests that on large enough scales,
 O\elem{VIII} emitters  trace the same underlying velocity field, 
characterized by linear, coherent flows, irrespective of their WHIM 
content. 
At smaller scales the iso-correlation contours 
appear to be elongated  along the line of sight which characterize
Fingers of God-type distortions.  In fact, these distortions appear to be
more prominent for WHIM-{\it dominated} emitters, while they seem to be 
absent in the {\it Total gas} case. 

To make our analysis more quantitative and assess the statistical significance 
of these distortions, we
have fitted the measured  $\xi(r_p,\pi)$ with an analytic model that 
accounts for both large-scale coherent motions and small-scales
by  expanding $\xi(r_p,\pi)$ in Legendre polynomials out to the quadrupole term
\citep{Kaiser87, Hamilton92, Cabre09}.
This model is fully characterized by two parameters: 
an ``elongation'' parameter $\sigma_{12}$, 
which quantifies the strength of Fingers of God-type distortions, 
and a ``compression'' parameter, $\beta$,
which quantifies the strength of linear-type distortions.
For a collisionless fluid, $\sigma_{12}$ measures the 
typical strength of incoherent motions while 
$\beta$ quantifies the amplitude of coherent 
motions, which is proportional to the growth of density fluctuations $f$:
$\beta \propto f(\Omega_m) \propto \Omega_m^{0.55}$, where 
 $\Omega_m$ is the mean mass density in units of critical density.
For O\elem{VIII} emitters such relations are not valid anymore. However,
one can still measure $\sigma_{12}$ and $\beta$, and compare them with 
the null case to detect the presence of either type of motions, or to compare the 
relative relevance of these motions among the different samples of
O\elem{VIII} emitters.

For this purpose we obtain the parameters $\beta$ and $\sigma_{12}$ 
by minimizing the $\chi^2$-like function
\begin{equation}
\chi^2=\frac{\sum_{ij}\left(y^m_{ij}(\beta,\sigma_{ij})-y_{ji}\right)^2}
	{\epsilon_{ij}^2},
\label{eq:chi2}
\end{equation}
where 
\begin{equation}
y=log\left(1+\xi(r_p,\pi)\right) ,
\label{eq:log}
\end{equation}
$\epsilon_{ij}$ is the Poisson error in the measured correlation function, 
the upper script $m$ indicates 
the model and the sum runs over all bins ($i,j$) in which we have 
sampled the ($r_p,\pi$) plane.

The distortion model provides a poor fit 
at small perpendicular separation, as demonstrated by the fact
that the reduced $\chi^2$ decreases significantly when one excludes 
pairs with  $r_p<6$~Mpc~h$^{-1}$ from the analysis.
This means that we cannot trust the value of $\sigma_{12}$ obtained from the 
fit and therefore we cannot estimate the statistical significance of
the Fingers of God-type distortions
that appear more prominent in the  WHIM-{\it dominated} emitters.

On the contrary, the model provides a good fit for  $r_p>6$~Mpc~h$^{-1}$ 
and the best fitting value of $\beta$ appears to be robust 
 since it does not change when including/excluding pairs 
with  $r_p<6$~Mpc~h$^{-1}$ from the fit or when we fix
$\sigma_{12}=0$.
In Tab.~\ref{table_bestfit} we list the best fit value of $\beta$ and its 1-$\sigma$ error
for the three types of emitters considered. We only consider the case in which we 
exclude pairs with $r_p<6$~Mpc~h$^{-1}$ since changing the cuts does not 
alter the result. For all types of emitters we measure a compression parameters
$\beta$ significantly above zero. This means that gas responsible for detectable 
O\elem{VIII} lines is typically found in regions that are moving coherently toward 
large density concentrations, as expected for O\elem{VIII} emitters that populate 
the outskirts of virialized structures rather than their central regions.
Considering the approximation involved in our error analysis and the uncertainties in the 
models of WHIM, we do not regard the differences among the values of 
$\beta$ for the different emitters as statistically significant.

\begin{table}
\begin{center}
\caption{Best-fit values of the compression parameter $\beta$
in the $\xi(r_p,\pi)$ at $r_p>6$~Mpc~h$^{-1}$
\label{table_bestfit}}
\begin{tabular}{|c|c|}
\hline\hline
Emitter type & $\beta$ \\
\hline
{\it Bright} WHIM      									&$0.179\pm0.023$  \\
{\it Total gas}	&$0.152\pm0.018$\\
WHIM-{\it dominated}		&$0.108\pm0.022$   \\
\hline
\end{tabular}
\end{center}
\end{table}

\section{Discussion and conclusions}
\label{conclusions}

In this paper we have addressed two different, but tightly related, issues: 
the feasibility of indirect detection of the WHIM through the angular 
correlation analysis of the DXB and the study and characterization 
of the spatial distribution of the WHIM {\it already detected} 
through its O\elem{VIII} X-ray line emission.

In this paper we have used  the WHIM model described by 
\citet{Branchini09} and \citet{Ursino10} based on 
the hydrodynamical simulation of \citet{Borgani04} 
with the metallicity-density relation taken from
Cen \& Ostriker (1999). 
Detectability and feasibility
estimates assume next-generation 
X-ray satellite missions. As a reference case we have
considered two instruments proposed in the framework
of  the proposed \Xenia mission.
For the angular correlation analysis of the DXB we have 
considered the HARI CCD-detector, while for the spatial correlation
study we have considered the CRIS spectrometer.

The main results of our analysis can be summarized as follows:
\begin{itemize}
	\item The  DXB autocorrelation signal in the soft X-ray range [0.38-0.65] keV 
	is mostly contributed by gas in hot and dense environments like
	clusters or groups. The WHIM contribution is comparatively smaller but 
	significantly different from zero. However, the intrinsic correlation signals of the  
	WHIM and {\it non}-WHIM  are similar, implying that the different contributions
	to the ACF of the DXB reflect the different SBs of the two components.
	No unique feature in the ACF of the WHIM can be used to unambiguously 
	infer its presence  from the measured ACF. The zero-crossing of the WHIM ACF 
	at $\sim5'$ indicates the typical angular scale of the WHIM in emission and 
	sets the minimum angular resolution required to next generation instruments aimed 
	at detecting the WHIM in emission.
	\item According to our WHIM model, the {\it non}-WHIM signal originates
	 from compact regions associated to virialized structures, while the fainter WHIM 
	emission traces the outskirts of these regions. One can then hope to isolate 
	the WHIM contribution by filtering out the signal associated to 
	virialized structures, typically associated to the brightest spots in the maps. 
	We found that the ACF measured after removing the $50\%$ brightest pixels from the map 
	is largely ($\sim75\%$) contributed by the WHIM. 
  More refined filtering procedures, in which one removes pixels associated to 
  known objects, would certainly improve the results. 	
	\item The presence of an isotropic and comparatively brighter 
	Galactic foreground component represents the main challenge to the 
	WHIM indirect detection. We have found that a successful 
	extraction of the WHIM ACF signal requires both an accurate 
	modeling of the Galactic foreground and a long exposure time. The 
	first requirement allows one to model the Galactic foreground as Poisson
	noise to be accounted for in the analysis. The second requirement allows to
	minimize shot noise errors. Our analysis suggests that exposure times of
	[0.1-1]~Ms are required to measure the WHIM ACF.  Uncertainties in the
	required  exposure time generously allow for uncertainties in the WHIM model. 
	Furthermore, it is possible to reduce the impact of the Galactic 
	foreground studying the ACF at energies below $\sim570$~eV (the 
	local O\elem{VII} triplet). This means focusing at emission lines at 
	redshift significantly different from zero, thus removing the strong 
	contribution of the Galactic O\elem{VII} and O\elem{VIII} foreground 
	(as partly done in \citealt{Galeazzi09}).
	\item The AGN contribution to the ACF is small and easily removed 
	by masking out pixels associated to resolved sources. 
         In a deep ($\sim1$~Ms) exposure one expects to identify 
         $\sim2000$ AGN in a $1^\circ\times1^\circ$ field. The corresponding 
         fraction of removed pixels is a few \% of the total.
         Unresolved AGN are basically uncorrelated and provides an isotropic
         contribution significantly smaller than that of the galactic foreground.
         	\item O\elem{VIII} emitters that can be detected in a 1~Ms observation show 
	no density evolution in the range $z\in[0,0.5]$. Deviations from the constant 
	density case reflect the increasing volume of the resolution element with 
	the redshift and the patchy appearance of the WHIM-{\it dominated} component.
	\item The two-point spatial correlation function of the O\elem{VIII} emitters
	can be measured with high statistical significance out to separations of $\sim 10$~Mpc~h$^{-1}$.
	Its amplitude and shape are close to those of optically selected galaxies, 
	suggesting that both types of objects trace the same underlying mass distribution.
	O\elem{VIII} emitters contributed by the WHIM are more correlated, as expected in
	the framework of the peak-biasing model since they are preferentially found in the 
	outskirts of large virialized structures.
	\item The correlation properties of O\elem{VIII} emitters do not 
	significantly evolve in the interval $z\in[0,0.5]$, despite the 
	appreciable differential evolution in the mass fraction of the 
	different gas phases. 
	\item Anisotropies in the clustering of O\elem{VIII} emitters induced by their peculiar 
	velocities can be detected in the measured two-point correlation function and used 
	to characterize their dynamical state. We have shown that the two-point correlation function
	of O\elem{VIII} emitters detected after an observation of 1~Ms will allow to unambiguously 
	detect such distortions which, according to our predictions, should be characteristic of
	infall motions toward virialized regions.
\end{itemize}

 Our analysis indicates that next generation instruments will be able to detect the WHIM contribution
 to ACF of the DXB, but it will require a rather long exposure of $\sim 1$ Ms. This prediction seems to 
 be at variance with the recent results of \citet{Galeazzi09} who found indirect evidence 
 for a WHIM-like  component in the angular correlation correlation properties of the DXB
diffuse signal in several  {\it XMM-Newton} fields with $t_{\it exp}$ in the range [50-600]~ks.
As we argued in Section~\ref{foreground}, the discrepancy is probably apparent since
our predictions are to be regarded as conservative for two reasons. 
First of all, our model overestimates the SB of the {\it non}-WHIM components hence
artificially increasing their contribution to the total ACF compared to that of the WHIM.
Second of all, we use a very simplistic cleaning procedure that is not optimized to extract
the WHIM signal. A better treatment of the  {\it non}-WHIM component, capable of bringing
theoretical prediction into agreement with currents DXB constraints, and a more effective cleaning 
procedure, that preferentially removes pixels associated to known objects, as 
\citet{Galeazzi09} did, will significantly reduce exposure time required to perform the 
angular correlation analysis. With this respect the \citet{Galeazzi09} results give us confidence
that next generation instruments will be able to carry out a very effective 
observational campaign for the indirect search of the WHIM within a reasonably
limited amount of time.

In fact, the comparison with \citet{Galeazzi09} results constitutes a new observational test
for our WHIM model. The ACF  measured by \citet{Galeazzi09}
(their Fig.~1) is in agreement with the one predicted by our 
model (Fig.~\ref{countclean_fgrem}) for angular separations larger than $\theta=2'$.
At smaller separation the mismatch simply reflects the smoothing effect of the instrumental PSF
which we did not try to model in our mock maps.

The fact that our WHIM model passes this further observational tests(the others being 
the O\elem{VI} and O\elem{VII} absorbers abundance and the DXB intensity) is certainly reassuring, 
but is no proof that our model is correct and its predictions robust. 
This concern, common to all WHIM models, has triggered a number of  works aimed at
testing the sensitivity of model predictions on crucial, ill-known processes of metal diffusion and 
stellar feedback \citep{Bertone10, Tornatore10, Shen010}. 
Focusing on emission, we have seen that the main inconsistency of our model
is the excess SB of the {\it non}-WHIM gas, whereas the soft X-ray emission of the 
WHIM meets the observational constraints. The natural question, then, is whether 
changing theoretical prescriptions for metal diffusion and galaxy feedback may 
help reducing the {\it non}-WHIM emission.
One possibility is  to invoke an AGN-driven rather than supernovae-driven feedback 
mechanism. \citet{Tornatore10} found that, on average,  AGN feedback increases the mass fraction 
and metallicity of the WHIM in regions with overdensities between a few and 10.
However, the bulk of detectable WHIM emission is produced in regions with higher overdensity,  
in which AGN feedback decreases both the mass fraction and metallicity of the gas.
The analysis of \citet{Bertone10} is perhaps more relevant to us, since it focuses on
the O\elem{VIII} emission line. They show that detectability predictions are quite 
robust to the different model prescription with the exception of the cooling function.
They show that using a metal-dependent cooling instead of a metal-independent scheme
(as in our model) decreases by up to an order of magnitude the SB of the O\elem{VIII} line
in the bright areas ($SB\ge 1$\phot), dominated by {\it non}-WHIM emission, but does not 
significantly alter the SB of the fainter areas, largely contributed  by WHIM emission.
Then, on a qualitative level, the use of a  metal-dependent cooling scheme, would
help relieving the problem of our models.

Assessing, in a quantitative way, the impact of these results on the possibility of 
detecting and studying the WHIM is beyond the scope of this paper. Here we have 
simply estimated their effect by modifying our WHIM model
in a way that effectively (but qualitatively!) mimics the impact of using a metal-dependent 
cooling scheme. For this purpose we have assumed an {\it ad-hoc} density-metallicity relation
such that the metallicity of the gas with WHIM density remains unchanged, while it is
decreased in high density region.
We stress the fact that this is an {\it ad-hoc} model that only serves for
illustrative purposes. 
The result of using this metallicity scheme is to decrease the mean SB 
in high density regions by $\sim30\%$ while the one of the WHIM remains the same.
The net effect is to reduce the SB contrast of the two phases although the relative 
statistics does not change appreciably, as indicated by the fact that 
in both the new and original metallicity schemes 50\% of the  pixels 
are dominated by WHIM  emission. In terms of correlation analysis, 
the angular ACF  of the WHIM remains the same but its statistical 
significance after the cleaning procedure increases significantly, as 
expected.

The results of this work and the robustness of the WHIM model predictions, suggest that a 
next-generation mission like \Xenia will be capable of extracting and characterizing
the WHIM contribution to the angular correlation 
properties DXB. We also expect to detect enough O\elem{VIII} (and O\elem{VII})
emitters to characterize the spatial correlation properties of the WHIM and its 
dynamical state. We confirm that, even with long exposure, next-generation
instruments will only be capable of studying the spatial distribution of the WHIM
in metal-rich, high density regions associated to the outskirts of large virialized structures, 
and of detecting the signature of its infall motion. Baryons tracing the large scale filaments 
of the cosmic web will remain largely undetected.

\section{Acknowledgments}
\label{ack}

We are grateful to Stefano Borgani for providing us the outputs of 
\citet{Borgani04} hydrodynamical simulation realized using the 
IBM-SP4 machine at the ``Consorzio Interuniversitario del Nord-Est 
per il Calcolo Elettronico'' (CINECA), with CPU time assigned thanks 
to an INAF-CINECA grant. This work has been supported by contract 
ASI-INAF I/088/06/0 TH-018. Eugenio Ursino acknowledges financial 
contribution from contract ASI-INAF I/088/06/0 WP 15300.


\label{lastpage}

\end{document}